\documentclass[twocolumn]{aastex63}

\shorttitle{Glitches due to history-dependent avalanches}
\shortauthors{Carlin et al.}

\usepackage{xcolor} 
\usepackage{booktabs}	
\usepackage{array}		
\usepackage[inline, shortlabels]{enumitem}	
\usepackage{multirow}   
\usepackage{dcolumn}
\usepackage{amsmath}
\usepackage{siunitx}

\newcommand{\td}{\textrm{d}}

\newcommand{\xc}{X_\textrm{c}}
\newcommand{\xmax}{X_{\rm th,\,max}}

\begin{document}

\title{Long-term statistics of pulsar glitches due to history-dependent avalanches}

\correspondingauthor{Julian B. Carlin}
\email{jcarlin@student.unimelb.edu.au}

\author[0000-0001-5694-0809]{Julian B. Carlin}
\affiliation{School of Physics, University of Melbourne, Parkville, VIC 3010, Australia}
\affiliation{OzGrav, University of Melbourne, Parkville, Victoria 3010, Australia}

\author{Andrew Melatos}
\affiliation{School of Physics, University of Melbourne, Parkville, VIC 3010, Australia}
\affiliation{OzGrav, University of Melbourne, Parkville, Victoria 3010, Australia}

\begin{abstract}
Stress accumulation-relaxation meta-models of pulsar glitches make precise, microphysics-agnostic predictions of long-term glitch statistics, which can be falsified by existing and future timing data. Previous meta-models assume that glitches are triggered by an avalanche process, e.g. involving superfluid vortices, and that the probability density function (PDF) of the avalanche sizes is history-independent and specified exogenously. Here a recipe is proposed to generate the avalanche sizes endogenously in a history-dependent manner, by tracking the thresholds of occupied vortex pinning sites as a function of time. Vortices unpin spasmodically from sites with thresholds below a global, time-dependent stress and repin at sites with thresholds above the global stress, imbuing the system with long-term memory. The meta-model predicts PDFs, auto- and cross-correlations for glitch sizes and waiting times, which are provisionally inconsistent with current observations, unlike some previous meta-models (e.g. state dependent Poisson process), whose predictions are consistent. The theoretical implications are intriguing, albeit uncertain, because history-dependent avalanches embody faithfully the popular, idealized understanding in the literature of how vortex unpinning operates as a driven, stochastic process. The meta-model predicts aftershocks, which occur with larger than average sizes and longer than average waiting times after the largest, system-resetting glitches. This prediction will be tested, once more data are generated by the next generation of pulsar timing campaigns.	
\end{abstract}


\section{Introduction} \label{sec:intro}

More than 500 rotational glitches have been detected in radio pulsars\footnote{Up-to-date online catalogues are maintained at the Jodrell Bank Centre of Astrophysics, \url{http://www.jb.man.ac.uk/pulsar/glitches.html} \citep{Espinoza2011}, and the Australian National Telescope Facility, \url{https://www.atnf.csiro.au/research/pulsar/psrcat/} \citep{Manchester2005}}. The sample is large enough that it is meaningful to disaggregate the data and study the long-term statistics of glitch activity in some individual objects with enough recorded glitches. Quantities of interest include waiting time and size cross-correlations, autocorrelations, and probability density functions (PDFs) \citep{Melatos2008, Espinoza2011, Melatos2018, Howitt2018, Fuentes2019, Carlin2019quasi, Carlin2019ac}.

Given measurements of the above quantities, one can falsify popular phenomenological meta-models describing glitch activity as a stress-relax process, wherein stress accumulates between glitches and relaxes partially at a glitch. A meta-model unifies a broad class of stochastic processes under its umbrella while remaining agnostic about the specific microphysics and the physical nature of the stress. It makes concrete, falsifiable predictions about long-term glitch statistics. The pay-offs from falsifiability are considerable and have driven work in this area recently. For example, existing data have been used to discriminate between the state-dependent Poisson (SDP) and Brownian stress accumulation (BSA) meta-models in six objects with high glitch activity \citep{Carlin2020}. In the SDP meta-model, glitches are triggered stochastically via a Poisson process with a stress-dependent rate, while the stress increases deterministically between glitches \citep{Warszawski2013, Fulgenzi2017}. In the BSA meta-model, glitches are triggered deterministically when the stress reaches a threshold, while the stress evolves stochastically via a Brownian process between glitches \citep{Carlin2020}. Both meta-models are consistent with various flavors of microphysics in the literature, e.g. superfluid vortex avalanches \citep{Anderson1975, Warszawski2011}, starquakes \citep{Larson2002, Middleditch2006}, and hydrodynamic instabilities \citep{Andersson2003, Mastrano2005, Glampedakis2009}, but they embody important physical differences in how the stress accumulates and when the relaxation events are triggered.

The roughly scale invariant glitch size PDFs observed in some pulsars suggest that the stress relaxes via an \emph{avalanche process}, i.e. a relaxation event triggered somewhere in the star propagates to adjacent regions via some knock-on mechanism \citep{Jensen1998, Melatos2008}. ``Chain reactions'' of this kind are observed in Gross-Pitaevskii simulations \citep{Warszawski2011, Warszawski2012, Warszawski2013, Lonnborn2019} and mean-field or N-body simulations \citep{Khomenko2018, Howitt2020} of superfluid vortex avalanches. They are typical of other systems cited as analogues of pulsar glitches in the literature, e.g. self-organized critical systems like sand piles, earthquakes, and magnetic fluxoids in type II superconductors \citep{Field1995, Jensen1998, Aschwanden2018}. In the meta-models published to date, a key input is the \emph{conditional distribution of avalanche sizes}, $\eta[\Delta X^{(n)}\,|\,X(t_{n}^-)]$, specifically the PDF of the spatially averaged stress released during the $n$-th avalanche, $\Delta X^{(n)}$, if the stress just prior to the avalanche is $X(t_{n}^-)$. There is no way to infer the functional form of $\eta[\Delta X^{(n)}\,|\,X(t_{n}^-)]$ uniquely from neutron star data, although one can get a rough idea from the observed glitch size distribution. Quantities of interest, such as size and waiting time PDFs, cross-correlations, and autocorrelations all depend jointly on the form of $\eta[\Delta X^{(n)}\,|\,X(t_{n}^-)]$, however, opening the door to falsifiability \citep{Carlin2019quasi, Carlin2019ac, Carlin2020}. 

In this paper, we generalize the successful SDP meta-model by adding a recipe to calculate $\eta[\Delta X^{(n)}\,|\,X(t_{n}^-)]$ endogenously instead of stipulating it exogenously by fiat. We refer to this generalization as the endogenous-$\eta$ meta-model. The recipe makes concrete the following key idea, which underpins the traditional picture of superfluid vortex avalanches in the literature: \begin{enumerate*}[(i)] \item the vortex pinning strength varies randomly from one location to the next within the star; \item the stress $\Delta X^{(n)}$ released in a glitch is proportional to the number of locations where the stress $X$ exceeds a threshold, when the glitch is triggered; and \item the unpinned vortices are reassigned randomly to new pinning locations after the glitch. \end{enumerate*} In other words, as time passes, the distribution of occupied pinning sites is continually revised in a history-dependent fashion, as the star spins down secularly and glitches occur stochastically. This idea is based on the ``coherent noise'' meta-model introduced by \citet{Sneppen1997} to describe sand piles \citep{Newman1996}, earthquakes \citep{Newman1996}, and biological extinctions \citep{Newman1996a}\footnote{The term ``coherent noise'' is not perfectly apt in the context of pulsar glitches, where the stress is coherent, while the noise, namely thermal creep, acts incoherently throughout the star. We retain the original terminology for consistency with previous literature \citep{Sneppen1997, Melatos2009}.}. It has been repurposed successfully as a pulsar glitch meta-model in previous work \citep{Melatos2009}. 

\citet{Haskell2016} investigated a meta-model that connects the ``snowplow'' mechanism for hydrodynamically triggered glitches \citep{Pizzochero2011} to simulations of vortex unpinning and re-pinning in a two-component fluid framework \citep{Haskell2014}. It differs from the meta-model considered here in two fundamental ways: \begin{enumerate*}[(i)] \item it unpins a random fraction of vortices at each avalanche event, independent of the past avalanche history; and \item it assumes event waiting times are exponentially distributed (i.e. the waiting times do not depend on the stress, as in the SDP  meta-model)\end{enumerate*}. However it does explicitly consider the mutual friction between the different fluid components, and shows that these hydrodynamic considerations produce deviations from power-law size distributions and exponential waiting time distributions.

We emphasize that we do not elect to analyze the endogenous-$\eta$ meta-model because we favor it over other options. True, it is reasonable from a physical standpoint, but so too are other options. The main motivation is that it formalizes the dominant (albeit idealized) idea in the literature about how vortex unpinning leads to glitches. We therefore quantify for the first time the long-term statistics predicted by the broad consensus behind what determines $\eta[\Delta X^{(n)}\,|\,X(t_{n}^-)]$ in the superfluid vortex avalanche picture. As we show below, the long-term statistics are somewhat inconsistent with observations to date, although the conclusion is not final; more data are needed. As a matter of fact the exogenous SDP meta-model, where $\eta[\Delta X^{(n)}\,|\,X(t_{n}^-)]$ is stipulated by fiat instead of being calculated self-consistently, seems to be more consistent with the data. It is unclear what this result implies more broadly, but it is intriguing and likely to inspire more investigations. It serves as a reminder of the value of falsification of physically-motivated meta-models.

The paper is structured as follows: in Section \ref{sec:sec2} we specify how $\Delta X^{(n)}$ is decided at each glitch. In Section \ref{sec:sdp} we connect this method of determining $\Delta X^{(n)}$ at a glitch to the SDP meta-model. In Section \ref{sec:obs} we explore the long-term statistical observables predicted by the endogenous-$\eta$ meta-model, and discuss how they compare to current glitch observations in Section \ref{sec:disc}. We conclude in Section \ref{sec:concl}. 

\section{Pinning threshold distribution} \label{sec:sec2}
We start by formalizing an idealized version of how the coherent stress mechanism may operate, which reflects standard ideas in the literature about stress accumulation and relaxation in pulsar glitches; see \citet{Haskell2015} for a recent review. As argued below, there are at least three reasons why the simple version may not be what is truly occurring: \begin{enumerate*}[(i)]
\item it introduces a cross-correlation between sizes and waiting times, which is largely absent from the data;
\item it struggles to generate exponentially distributed waiting times, which are seen in some pulsars; and
\item it assumes a spatially uniform stress distribution which is qualitatively different to the spatially correlated stress distribution observed in Gross-Pitaevskii \citep{Warszawski2011, Lonnborn2019} and N-body simulations \citep{Howitt2020} of superfluid vortex avalanches.
\end{enumerate*}
Nonetheless it is important to study the simple version first, partly to falsify it if possible, and partly because it highlights the key idea that the stress distribution is history-dependent. As in previous papers, we develop the meta-model in the context of superfluid vortex avalanches for concreteness. Adapting it to the context of other microphysics, e.g. starquakes, is possible but outside the scope of this paper \citep{Newman1996}.

\subsection{Standard picture}
To fix ideas, suppose that the region of the stellar interior where glitch activity occurs is divided into ``sites''. In the vortex avalanche picture, for example, the sites are nucleons or interstices in the nuclear lattice, where a superfluid vortex may pin \citep{Jones1998, Donati2006, Avogadro2007}; in the starquake picture, the sites are segments of a fault or other tectonic element of the rigid crust \citep{Ruderman1998, Middleditch2006, Chugunov2010}. At an arbitrary site located at $\mathbf{r}$, there is a local stress $X(\mathbf{r}, t)$ at time $t$, which evolves in response to the global driver and local relaxation physics. In the vortex avalanche picture, $X(\mathbf{r}, t)$ is proportional to the Magnus force; in the starquake picture, $X(\mathbf{r}, t)$ is proportional to the elastic stress (or equivalently the strain in the linear regime). 

Let $X_{\rm th}(\mathbf{r})$ denote the stress threshold at $\mathbf{r}$, which is assumed to be constant on the time-scales of interest (see below). Whenever the stress satisfies $X(\mathbf{r}, t) \geq X_{\rm th}(\mathbf{r})$, and a glitch is triggered, the stress relaxes locally and is redistributed to nearby sites. In the vortex avalanche picture, a vortex at a site unpins when it escapes the nuclear pinning potential; in the starquake picture, the crustal lattice fails locally when the breaking strain is exceeded. On the other hand, if the stress satisfies $X(\mathbf{r}, t) < X_{\rm th}(\mathbf{r})$, it remains supported stably at $\mathbf{r}$ during a glitch. A group of contiguous sites with $X(\mathbf{r}, t) < X_{\rm th}(\mathbf{r})$ form a stress reservoir or capacitive domain \citep{Alpar1996}. Note that $X(\mathbf{r}, t) > X_{\rm th}(\mathbf{r})$ can be supported metastably for some time before a glitch is triggered.

The coarse-grained stress threshold is assumed to be spatially uniform, because the length-scale of vortex avalanches or starquakes (inferred from the size of observed glitches) is smaller than the length-scale over which the nuclear lattice is stratified (typically $\sim10^2\,$m) \citep{Chamel2008}. However the fine-grained stress threshold varies randomly from one site to the next, due to defects and microscopic compositional gradients in the lattice. 

\subsection{Available versus occupied sites} \label{sec:g_intro}
When describing the pinning thresholds statistically, it is essential to distinguish between their ``available'' distribution (defined without reference to any vortices) and their ``occupied'' distribution (defined with reference only to those pinning sites where a vortex is pinned). These are not the same in general. The available distribution is governed by the physics of the nuclear lattice. The occupied distribution is governed by a combination of the available distribution and the history-dependent vortex dynamics; vortices may congregate preferentially in deeper pinning potentials, for example.

Let $\phi(X_{\rm th})$ denote the PDF of the thresholds at all available pinning sites. We assume that $\phi(X_{\rm th})$ is constant in time, because the nuclear properties of the star evolve slowly (e.g. on the cooling time-scale $\sim10^5\,$yr) compared to the inter-glitch waiting time and the length of glitch monitoring campaigns to date. Let $\xmax = {\rm max}_{X_{\rm th}} X_{\rm th}$ denote the maximum pinning threshold in the star. In the superfluid vortex picture, for example, $\xmax$ corresponds to the nuclear site with the deepest pinning potential; if a uniform stress satisfying $X \geq \xmax$ is applied to the system, every vortex unpins when a glitch is triggered. The functional form of $\phi(X_{\rm th})$ is unknown from first principles, but theoretical calculations suggest that it is broad, with standard deviation comparable to the mean \citep{Jones1998, Donati2006, Avogadro2007}, and $\xmax$ is certain to be finite as stipulated by quantum mechanics. In this work we assume $\phi(X_{\rm th})$ has compact support on the interval $0 = X_{\rm th,\,min} \leq X_{\rm th} \leq \xmax$ for simplicity, but the results carry over straightforwardly to the case $X_{\rm th,\,min}\neq 0$ without changing qualitatively. We emphasize that $\xmax$ need not equal $\xc$, the critical stress at which a glitch is certain in either the SDP or BSA meta-models. This subtle point is discussed further in Section \ref{sec:qual}.

What is the PDF $g(X_{\rm th}, t)$ of the stress thresholds at pinning sites actually occupied by vortices at time $t$? In the absence of a global driver, we have $g(X_{\rm th},t) = \phi(X_{\rm th})$ in the limit $t\rightarrow \infty$, after initial transients die away. In the presence of a persistent global driver, however, $g(X_{\rm th}, t)$ depends on time and does not equal $\phi(X_{\rm th})$ for any $t$ in general (except perhaps $t = 0$, or in the special situation where all vortices unpin at once, resetting the system back to $g(X_{\rm th}, t) = \phi(X_{\rm th})$, as described in Section \ref{sec:unpin}). Whenever a glitch occurs, vortices unpin from relatively shallow pinning sites and repin randomly, changing the relative occupation of sites with lower and higher $X_{\rm th}$. Thus $g(X_{\rm th}, t)$ is not only time- and history-dependent but also stochastic; it depends on the random sequence of glitch sizes and waiting times up to the instant $t$.

How $g(X_{\rm th}, t)$ evolves depends, among other things, on whether one treats the system as spatially uniform or nonuniform. In self-organized critical systems like sand piles, for example, long-range spatial correlations exist between the stress at different locations $\mathbf{r}$ and $\mathbf{r'}$, with $|\langle X(\mathbf{r}, t) X(\mathbf{r'}, t)\rangle - \langle X(\mathbf{r}, t)\rangle \langle X(\mathbf{r'}, t)\rangle | \propto |\mathbf{r} - \mathbf{r'} |^{-a}$ and $a > 0$ typically \citep{Jensen1998, Aschwanden2018}. The system self-organizes through spatial gradients to produce scale-invariant dynamics (e.g. power-law avalanche size PDFs). A strong case can be made, through Gross-Pitaevskii and N-body simulations, that superfluid vortex avalanches or starquakes in a neutron star are self-organized critical systems too \citep{Melatos2008, Haskell2015, Howitt2020}. However, the theoretical treatment of a far-from-equilibrium system with correlated spatial gradients is a notoriously challenging (and unsolved) problem in statistical mechanics \citep{Jensen1998}. In this paper, therefore, we derive an equation of motion for $g(X_{\rm th}, t)$ under the assumption that every pinning site in the star experiences the same, spatially-averaged stress, $X(\mathbf{r},t) = X(t)$. This mean-field approximation \citep{Fulgenzi2017, Khomenko2018} has been employed successfully in the analysis of pulsar glitch observational data on size and waiting time PDFs, cross-correlations, and autocorrelations in the context of the SDP and BSA meta-models \citep{Fulgenzi2017, Melatos2018, Carlin2019quasi, Carlin2019ac, Carlin2020, Melatos2019} as well as in previous work on ``coherent noise'' models of pulsar glitches [see footnote 2 and \citet{Melatos2009}].

\subsection{Unpinning and repinning} \label{sec:unpin}
Consider two consecutive glitches that occur at times $t_n$ and $t_{n+1}$. Let $t_n^\pm$ denote the instants infinitesimally after ($+$) and before ($-$) the event at $t = t_n$\footnote{It is important to distinguish $t_n$ from $t_n^\pm$, because the glitch is assumed to occur instantaneously, as in previous work.}. During the interval $t_n^+ \leq t \leq t_{n+1}^-$, the vortices are pinned, so the occupied sites do not change, and neither does $g(X_{\rm th},t)$. That is, we have $g(X_{\rm th}, t) = g(X_{\rm th}, t_n^+)$ for $t_n^+ \leq t \leq t_{n+1}^-$. We also have $g(X_{\rm th}, t) = 0$ for $X_{\rm th} \leq X(t_n^+)$, because vortices repin exclusively at sites with $X_{\rm th} > X(t_n^+)$, when the avalanche at $t = t_n$ occurs; they cannot get stuck at sites where the stress exceeds the local threshold, while they are moving freely. Simultaneously, $X(t)$ evolves during the interval $t_n^+ \leq t \leq t_{n+1}^-$ (deterministically in the SDP process, stochastically in the BSA process). By the time the instant $t = t_{n+1}$ is reached, we have $X(t_{n+1}^-) \geq X_{\rm th}$ at many sites; the corresponding vortices are pinned metastably and are ready to unpin when provoked by some minuscule statistical (e.g. thermal) fluctuation. When they do unpin, they are assumed to do so instantaneously at $t = t_{n+1}$. The instantaneous approximation is justified amply by Gross-Pitaevskii simulations, where vortex avalanches are observed to occur on a time-scale shorter than one rotation period, and by high-time-resolution radio timing observations, which reveal that the spin up during a glitch takes less than $\sim30\,$s \citep{McCulloch1990, Dodson2002, Palfreyman2018, Ashton2019}. Both these time-scales are much shorter than the typical inter-glitch waiting time of weeks to months.

How many vortices unpin at $t=t_{n+1}$? In the mean-field approximation, the answer is all of them pinned in sites that satisfy $X_{\rm th} \leq X(t_{n+1}^-)$, which represent a fraction, $F$, of the total, with 
\begin{eqnarray}
F =& \int_0^{X(t_{n+1}^-)} \td X_{\rm th}' g(X_{\rm th}', t_{n+1}^-) \\
  =& \int_{X(t_n^+)}^{X(t_{n+1}^-)} \td X_{\rm th}' g(X_{\rm th}', t_{n}^+)\ . \label{eq:f}
\end{eqnarray}
The unpinned vortices move radially outward by a distance $\Delta r$ comparable to one inter-vortex (Feynman) separation $\lambda_F$ before repinning (as seen in Gross-Pitaevskii and N-body simulations), reducing the angular momentum of the superfluid in proportion to $\Delta r$ and their number. By angular momentum conservation, the crust spins up, and the crust-superfluid angular velocity lag (i.e. the spatially averaged stress) decreases. That is, we have
\begin{eqnarray}
X(t_{n+1}^+) - X(t_{n+1}^-) = -KF\ , \label{eq:delx}
\end{eqnarray}
where $K > 0$ is a constant measured in units of stress. To ensure $X(t) \geq 0$ we require $K \leq \xmax$. 

In the vortex avalanche picture, where $X$ is the crust-core differential angular velocity, angular momentum conservation implies \citep{Melatos2015, Fulgenzi2017}
\begin{eqnarray}
KF &= \frac{2\pi(I_{\rm C} + I_{\rm S}) \Delta \nu}{I_{\rm S}} \label{eq:kf}
\end{eqnarray}
or equivalently
\begin{eqnarray}
K = \frac{2\pi(I_{\rm C} + I_{\rm S})}{I_{\rm C}} \frac{\nu\Delta r}{R} \ , \label{eq:kr}
\end{eqnarray}
where $\nu$ is the rotational frequency of the rigid crust (with $X \ll \nu$), $\Delta \nu$ is the increase in $\nu$ at the glitch, $I_{\rm S}$ and $I_{\rm C}$ are moments of inertia of the superfluid and crust respectively, and $R$ is the neutron star radius. For $\Delta r \sim \lambda_F = \num{4.1e-2} (\nu/10\,{\rm Hz})^{-1/2}\,$cm, we estimate 
\begin{eqnarray}
K \sim 10^{-7} \left(\frac{I_{\rm C} + I_{\rm S}}{I_{\rm C}}\right)\left(\frac{\nu}{10\,{\rm Hz}}\right)^{1/2} \left(\frac{R}{10\,{\rm km}}\right)^{-1}\, {\rm Hz}.
\end{eqnarray}
The factor $(I_{\rm C} + I_{\rm S}) / I_{\rm C}$ amounts to $\sim10^{2}$ if the crust is a thin crystalline lattice while the rest of the star is composed of a superfluid \citep{Andersson2012, Hooker2015} or $\sim 1$ if the crust is coupled tightly to most of the neutrons and protons throughout the star, e.g. by pinning between neutron vortices and magnetic fluxoids of the superfluid \citep{Srinivasan1990, Link1999a, Lyne2000a, Espinoza2011}. As discussed in Section 2.5 of \citet{Carlin2020} the coupling between the rotation of the superfluid and the crust may be imperfect \citep{Pizzochero2020}. In light of the uncertainty, we allow $K$ to remain a free parameter in the meta-model.

The unpinned vortices repin instantaneously. The number of pinning sites per unit area is $\sim10^{20}$ times the number of vortices per unit area, so there is no reason for a vortex to pin preferentially to a shallower or deeper pinning site, as it circulates freely during an avalanche. Hence the probability that a vortex repins at a site with stress threshold $X_{\rm th}$ is simply proportional to the number of such sites present in the crustal lattice, i.e. the fraction of the total sites with threshold $X_{\rm th}$. Therefore the distribution of thresholds at which the vortices repin satisfies $g_{\rm repin}(X_{\rm th}, t_{n+1}^+) \propto \phi(X_{\rm th})$\footnote{In other words, as there many more pinning sites than vortices, the distribution of thresholds of pinning sites at which a free vortex may repin is equal to the distribution of thresholds of pinning sites in general. A free vortex is not biased towards pinning sites with higher or lower thresholds. It pins indiscriminately to any site, whose threshold exceeds the global stress \citep{Haskell2016a}.}. Of course, freely moving vortices cannot repin at a site whose threshold is lower than the global stress, as discussed above, so we have
\begin{eqnarray}
g_{\rm repin}(X_{\rm th}, t_{n+1}^+) = A\, \phi(X_{\rm th})\, H[X_{\rm th} - X(t_{n+1}^+)]\ , \label{eq:grepin}
\end{eqnarray}
where $H(...)$ denotes the Heaviside function. The constant $A$ is determined by normalization with respect to the unpinned fraction and is given by
\begin{eqnarray}
A = \left[\int_{X(t_{n+1}^+)}^{\xmax} \td X_{\rm th}' \phi(X_{\rm th}') \right]^{-1} F\ . \label{eq:anorm}
\end{eqnarray}

\subsection{Equation of motion for the distribution of occupied pinning sites} \label{sec:eom}

\begin{figure*}[t]
\centering
\includegraphics[width=\linewidth]{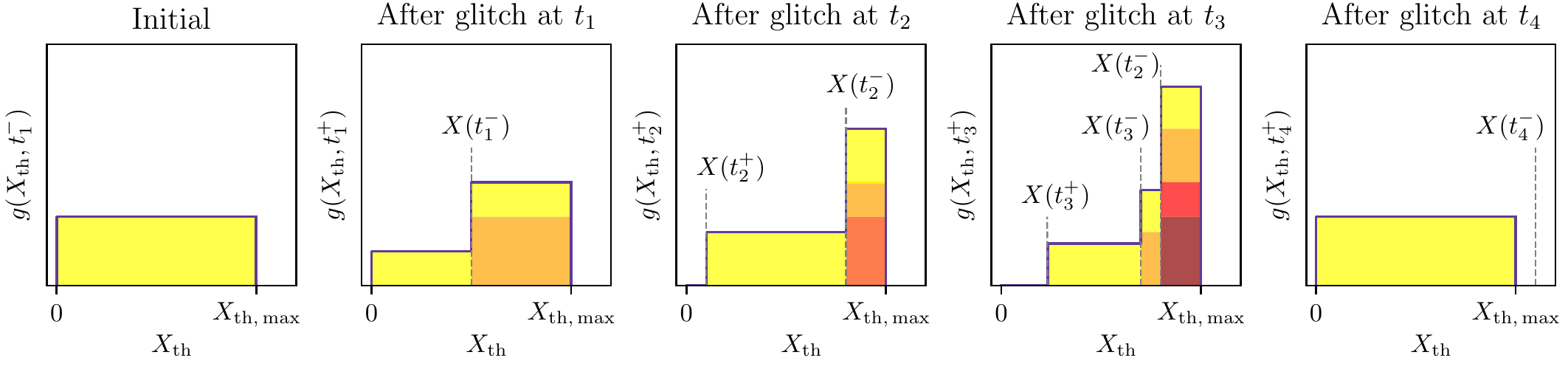}
\caption{A toy example illustrating schematically the evolution of $g(X_{\rm th}, t)$ during a sequence of four glitches, with $X(t_1^+) < X(t_3^+) < X(t_2^+) < \xmax < X(t_4^-)$, where $g(X_{\rm th}, t_1^-) = \phi(X_{\rm th})$ is the uniform distribution between $0 \leq X_{\rm th} \leq \xmax$. The colors correspond to how many glitches occur in the time elapsed since vortices unpin from pinning sites with a certain threshold. Darker colors indicate that the vortices have stayed pinned during more events. \label{fig:g_illust}}
\end{figure*}

We are now in a position to combine the results on unpinning and repinning in Section \ref{sec:unpin} to write down the equation of motion for $g(X_{\rm th},t)$. We remind the reader that $g(X_{\rm th},t)$ does not evolve at all during the inter-glitch interval $t_n^+ \leq t \leq t_{n+1}^-$, while vortices are pinned. Its evolution is described completely, in the above approximation, by discontinuous adjustments at each glitch. That is, at $t = t_{n+1}$, the following things happen to the stress threshold PDF: \begin{enumerate}[(i)]
\item we start with $g(X_{\rm th}, t_{n+1}^-) = g(X_{\rm th},t_n^+)$, which is nonzero for $X_{\rm th} > X(t_n^+)$; 
\item we calculate the change in stress $-KF$ in equation \eqref{eq:delx} by integrating $g(X_{\rm th},t_n^+)$ according to equation \eqref{eq:f}; 
\item we set $g(X_{\rm th},t_{n+1}^-)$ to zero in the range $X(t_n^+) \leq X_{\rm th} \leq X(t_{n+1}^-)$, because the vortices therein unpin; 
\item we reassign the unpinned vortices to sites with thresholds in the range $X(t_{n+1}^+) < X_{\rm th} \leq \xmax$ according to equations \eqref{eq:grepin} and \eqref{eq:anorm}. \end{enumerate} Putting steps (i)--(iv) together yields
\begin{widetext}
\begin{eqnarray}
g(X_{\rm th}, t_{n+1}^+) =&~ g(X_{\rm th}, t_{n+1}^-) H[X_{\rm th} - X(t_{n+1}^-)] + \left[\int_{X(t_{n+1}^+)}^{\xmax} \td X_{\rm th}' \phi(X_{\rm th}') \right]^{-1} F\,\phi(X_{\rm th}) H[X_{\rm th} - X(t_{n+1}^+)] \label{eq:geom}
\end{eqnarray}
\end{widetext}

It is easy to check that this normalizes correctly with $1 = \int_0^{\xmax}\td X_{\rm th}' g(X_{\rm th}', t_{n+1}^+)$. Note that $g(X_{\rm th}, t_{n+1}^+)$ depends on $g(X_{\rm th}, t_n^+)$ implicitly through $F$. It also depends on $X(t_{n+1}^-)$ and $X(t_n^+)$ independently, because the evolution of $X(t)$ during the interval $t_n^+ \leq t \leq t_{n+1}^-$ is controlled by the stress accumulation process, which does not depend on the unpinning and repinning physics.

Equation \eqref{eq:geom} resembles, but is not the same as, the equations of motion in \citet{Sneppen1997} and \citet{Melatos2009}. The foregoing papers treat unpinning and repinning similarly, i.e. by nullifying the unpinned portion of $g(X_{\rm th},t)$ and reassigning it elsewhere $\propto \phi(X_{\rm th})$\footnote{They also include an optional thermal unpinning process independent of $X_{\rm th}$, which is not essential and is omitted in this paper.}. However neither paper evolves $g(X_{\rm th},t)$. It is assumed that $g(X_{\rm th},t)$ converges rapidly to its steady-state form, $g(X_{\rm th},t) = g(X_{\rm th})$, i.e. the system establishes detailed balance between unpinning and repinning at each individual value of $X_{\rm th}$ [see equations (3) and (4) in \citet{Melatos2009}]. This approach is perfectly defensible, when the goal is to calculate the statistically stationary size and waiting time PDFs, but it does not contain enough information to study the cross- and autocorrelations we are interested in here \citep{Melatos2018, Carlin2019quasi, Carlin2019ac}.

To build intuition, Figure \ref{fig:g_illust} illustrates schematically the evolution of $g(X_{\rm th}, t)$ over the course of four hypothetical glitches at $t_1$, $t_2$, $t_3$, and $t_4$. For simplicity we choose $\phi(X_{\rm th})$ to be uniform in the range $0 \leq X_{\rm th} \leq \xmax$, with $K = \xmax$. The system starts with $g(X_{\rm th}, t_1^-) = \phi(X_{\rm th})$. In this particular realization we choose $X(t_1^+) < X(t_3^+) < X(t_2^+) < \xmax < X(t_4^-)$ for illustrative purposes. The first glitch, at $t_1$, unpins half of the vortices, and they are reassigned according to equation \eqref{eq:grepin}. The second glitch, at $t_2$, unpins more than half of the vortices, but due to the previous glitch we have $X(t_2^+) \neq 0$, as $g(X_{\rm th}, t)$ is no longer uniform. The third glitch, at $t_3$, occurs with $X(t_3^-) < X(t_2^-)$, demonstrating that the memory of the previous stress in the system is imprinted on $g(X_{\rm th}, t)$, i.e. $g(X_{\rm th}, t_3^+)$ depends on $X(t_2^-)$, not just $X(t_3^{\pm})$. The fourth glitch, at $t_4$, resets the system back to $\phi(X_{\rm th})$ because we have $X(t_4^-) > \xmax$, and so all vortices unpin. 

Note that there is a complex feedback loop at play which relates the glitch sizes and waiting times to $g(X_{\rm th}, t)$. If chance produces a long sequence of frequent glitches, i.e. the stress does not increase much before another glitch is triggered, vortices pile up at larger values of $X_{\rm th}$ near $\xmax$. Then, when there is a long delay, which gives the stress time to reach $X \approx \xmax$, the vortex pile pinned at sites with $X_{\rm th} \approx \xmax$ unpins all at once to produce a relatively large glitch resetting the system. By contrast, a similarly long delay produces a smaller glitch, if it is preceded by a glitch sequence which does not pile up vortices at $X_{\rm th} \approx \xmax$.

\section{State-dependent Poisson process} \label{sec:sdp}
The above recipe for updating $g(X_{\rm th},t)$ must be combined with a compatible recipe for choosing the waiting times between glitches. Two approaches have been explored previously, in the SDP \citep{Fulgenzi2017, Carlin2019quasi} and BSA meta-models \citep{Carlin2020}. In the BSA meta-model the stress accumulates stochastically to a fixed threshold, whereupon a glitch is triggered deterministically. In the SDP meta-model glitches are triggered probabilistically, at an instantaneous rate which depends on the stress in the system. If $X(t_n^-) = \xc$ is identical for each glitch as in the BSA meta-model, the system does not produce glitches of different sizes, as the same fraction of vortices would unpin every time, and repopulate the same distribution of pinning sites. To avoid this trivial behavior, which is inconsistent with pulsar data, we henceforth adopt the SDP meta-model to pick waiting times, and therefore determine $X(t_{n+1}^-)$, given $X(t_n^+)$.

We position the SDP and related meta-models in the broader context of point processes and time series modeling in Appendix \ref{sec:timeseries}.

\subsection{Equation of motion for the stress} \label{sec:xeom}
The stress, $X(t)$, in the star evolves according to 
\begin{eqnarray}
X(t) &= X(0) + t + \sum_{n=1}^{N(t)} \Delta X^{(n)}\ , \label{eq:xeom}
\end{eqnarray}
where $X$ and $t$ are here and henceforth expressed in dimensionless units of $\xc$ (the critical stress at which a glitch becomes certain) and $\xc I_{\rm C} / N_{\rm em}$ respectively, where $N_{\rm em}$ is the electromagnetic torque acting on the crust, and $X(0)$ is an arbitrary initial condition. The number of glitches up to time $t$, $N(t)$, is a random variable, implicitly determined by the waiting times between each glitch. The size $\Delta X^{(n)}$ of the $n$-th relaxation event is a deterministic function of the fluctuating PDF $g(X_{\rm th}, t)$. The recipe for determining $\Delta X^{(n)} = -KF$ is outlined in Section \ref{sec:unpin}, specifically equations \eqref{eq:f} and \eqref{eq:delx}.

One key assumption of the SDP meta-model is that the instantaneous glitch rate, $\lambda(t)$, is a function of the spatially-averaged stress, $X(t)$. We assume $\lambda[X(t)]$ grows monotonically between glitches according to
\begin{eqnarray}
\lambda[X(t)] &= \frac{\alpha}{1 - X(t)}\ ,
\end{eqnarray}
where 
\begin{eqnarray}
\alpha = \frac{I_{\rm C} \xc \lambda_0}{N_{\rm em}}
\end{eqnarray}
is a dimensionless control parameter, and $\lambda_0$ is a reference rate defined as $\lambda_0 = \lambda(1/2)/2$. The exact functional form of $\lambda[X(t)]$ does not significantly change the long-term dynamics, as long as the rate diverges as $X \rightarrow \xc$ \citep{Fulgenzi2017, Carlin2019quasi}. 

The PDF of waiting times after the $n$-th glitch is \citep{Cox1955, Fulgenzi2017}
\begin{eqnarray}
p[\Delta t\,|\,X(t_n^+)] = &~\lambda&[X(t_n^+) + \Delta t] \label{eq:pdelt}\\
&\times&\exp \left\{-\int_{t_n^+}^{t_n^++\Delta t}\td t' \lambda[X(t_n^+) + t'] \right\}. \nonumber
\end{eqnarray}

\subsection{Monte Carlo simulations} \label{sec:mcsteps}
The evolution of $X(t)$ and $g(X_{\rm th}, t)$ is jointly modeled with a simple Monte Carlo automaton.
\begin{enumerate}
\item Initialize the system at $t=0$ with $g(X_{\rm th}, 0) = \phi(X_{\rm th})$, and $X = X(0)$.
\item Pick a random $\Delta t$ from equation \eqref{eq:pdelt}, given the current stress $X$.
\item Update the stress to $X + \Delta t$ to account for the deterministic evolution up to the glitch.
\item Evaluate $\Delta X = -KF$ deterministically via equations \eqref{eq:f} and \eqref{eq:delx}, given $g(X_{\rm th}, t)$ and $X$.
\item Update $g(X_{\rm th}, t)$ according to equation \eqref{eq:geom}.
\item Update $X$ by adding $\Delta X$ according to equation \eqref{eq:xeom}.
\item Repeat from step 2.
\end{enumerate}
Random numbers for step 2 are picked using the standard inverse cumulative algorithm \citep{Press2007}. As $\phi(X_{\rm th})$ is chosen to be a uniform distribution between $0\leq X_{\rm th} \leq \xmax$, $g(X_{\rm th}, t)$ is a piecewise-constant function. Hence, we efficiently update it by storing the heights at change-points $X_{\rm th}$ in memory, as opposed to sampling $g(X_{\rm th}, t)$ on a grid of $X_{\rm th}$ values. 

\subsection{Qualitative results} \label{sec:qual}

\begin{figure}[t]
\centering
\includegraphics[width=0.95\linewidth]{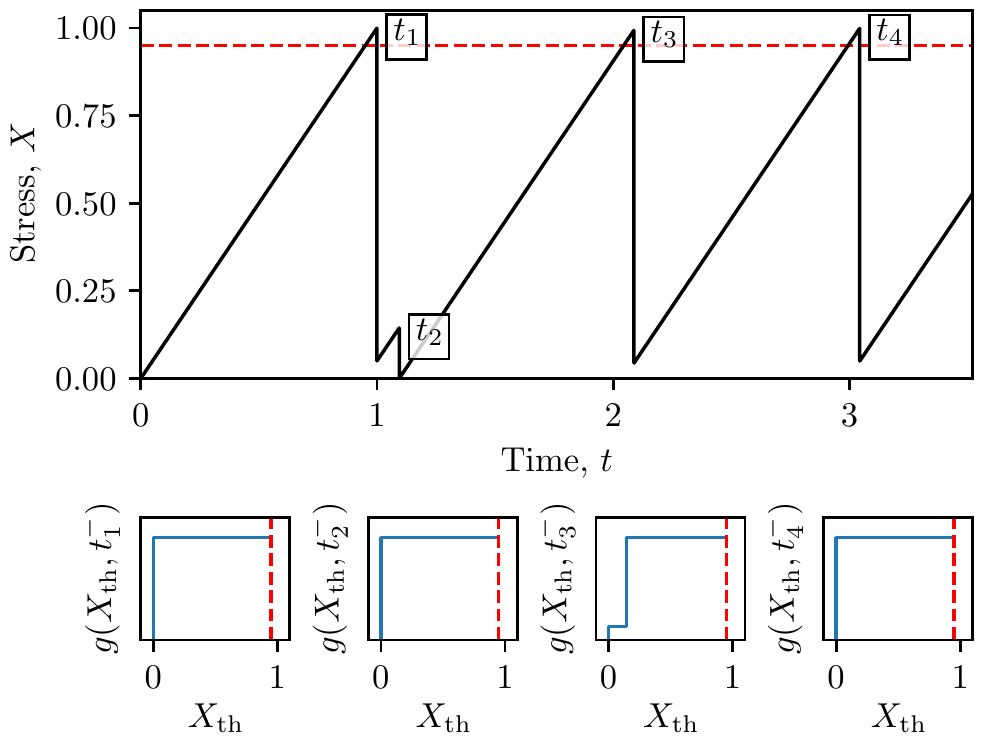}
\caption{Evolution of $X$ (main panel) and $g(X_{\rm th}, t)$ (sub-panels) across four glitches at $t_1, \dots, t_4$, generated via the automaton outlined in Section \ref{sec:mcsteps}. Parameters: $\alpha=0.2$, $\xmax=0.95$ (indicated in both the main panel and sub-panels with a red dashed line), $K=\xmax$. $X$ and $X_{\rm th}$ are in units of $\xc$ and $t$ is in units of $\xc I_{\rm C} / N_{\rm em}$.\label{fig:lowa_4}}
\end{figure}

\begin{figure}
\centering
\includegraphics[width=0.95\linewidth]{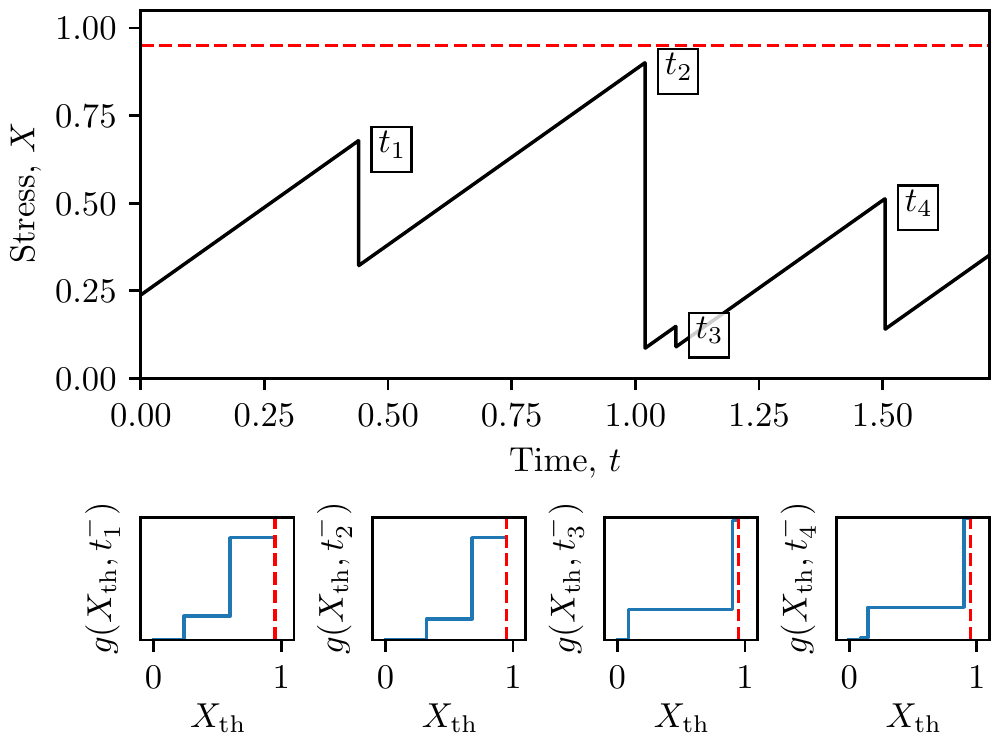}
\caption{As in Figure \ref{fig:lowa_4} but with $\alpha=1$. \label{fig:mida_4}}
\end{figure}

\begin{figure}
\centering
\includegraphics[width=0.95\linewidth]{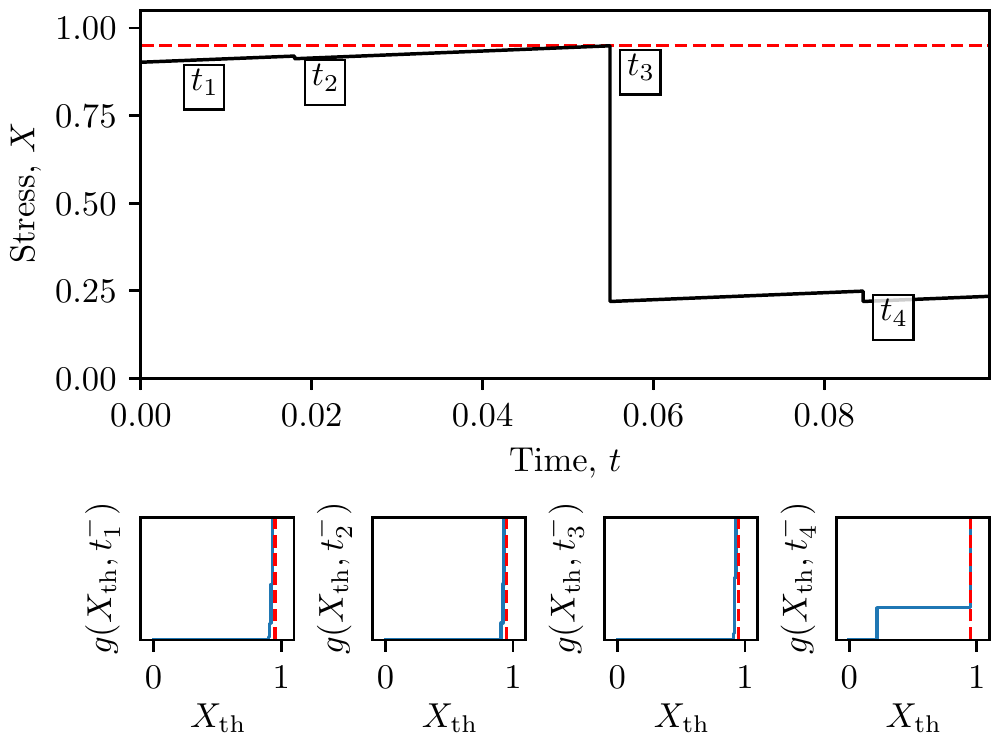}
\caption{As in Figure \ref{fig:lowa_4} but with $\alpha=10$.\label{fig:higha_4}}
\end{figure}

There are three control parameters in the dimensionless meta-model: $\alpha$, $\xmax$, and $K$. We fix $\phi(X_{\rm th})$ to be a uniform distribution between $0$ and $\xmax$. For simplicity, we assume $K=\xmax$ (its maximal value) for the rest of this work. Appendix \ref{sec:lowk} contains a brief exploration of the impact $K < \xmax$ has on long-term observables. Figures \ref{fig:lowa_4}--\ref{fig:higha_4} illustrate the qualitative impact of $\alpha$ on the evolution of $X$ and $g(X_{\rm th}, t)$, given $\xmax = 0.95\xc$. Figure \ref{fig:lowa_4} shows that at low values $\alpha \lesssim 0.5$, longer waiting times are more likely, and the stress often exceeds $\xmax$ before each glitch, resetting the system. Therefore one finds $g(X_{\rm th}, t) \approx \phi(X_{\rm th})$, i.e. there is little long-term memory in the system. In Figure \ref{fig:higha_4} high values $\alpha \gtrsim 5$ produce shorter waiting times. Thus long sequences of small glitches unfold before rare, large events reset the system once the stress finally accumulates to $X \gtrsim \xmax$. Figure \ref{fig:mida_4} shows that for intermediate values of $\alpha$ between the above two extremes the behavior of both the stress $X(t)$ and the evolution of $g(X_{\rm th}, t)$ is complex. 

\begin{figure}[t]
\centering
\includegraphics[width=\linewidth]{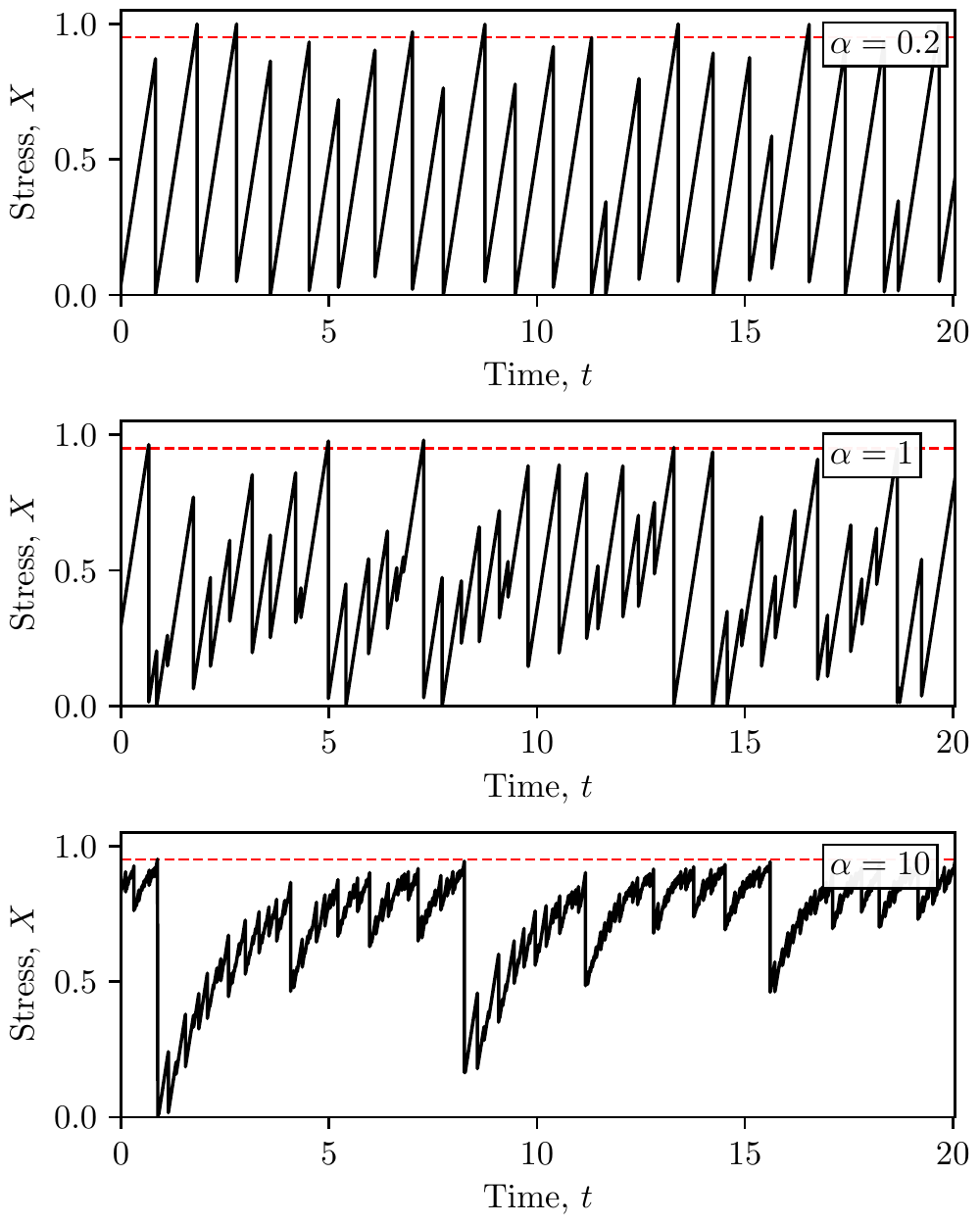}
\caption{Long-term qualitative behavior of $X$ for three different values of $\alpha = 0.2$, 1, and 10 in the top, middle and bottom panel respectively. Parameters: $\xmax=0.95$ (indicated with a red dashed line in each panel), $K=\xmax$. \label{fig:longterm}}
\end{figure}

The qualitative long-term behavior of $X$ is shown in Figure \ref{fig:longterm}. In the top panel with $\alpha=0.2$, the stress often exceeds $\xmax$ due to the long waiting times between glitches, resetting the system and keeping the average stress around $X \sim 0.5$. In the middle panel with $\alpha=1$, the average stress climbs stochastically until finally a longer than average waiting time allows $X > \xmax$, resetting the system. In the bottom panel with $\alpha=10$, waiting times are short and the stress builds asymptotically towards $X = \xmax$. For $\alpha \gtrsim 1$ the automaton takes some time to stabilize such that the memory of the arbitrary initial conditions $X(t=0) = 0$ and $g(X_{\rm th}, t=0) = \phi(X_{\rm th})$ is lost. When calculating long-term statistics predicted by the meta-model in Section \ref{sec:obs} we throw away the first $\lfloor100 \alpha\rfloor$ glitches generated.


\begin{figure}
\centering
\includegraphics[width=0.95\linewidth]{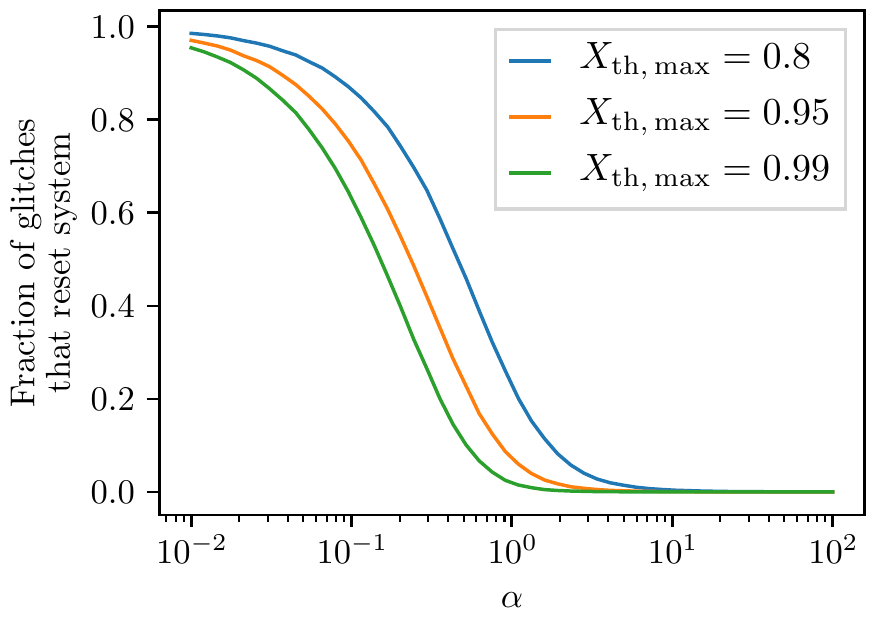}
\caption{Fraction of glitches that reset the system as a function of $\alpha$ for three different values of $\xmax$. Parameters: $N=10^5$ glitches generated at 50 logarithmically spaced values of $\alpha$ for each value of $\xmax$, with $K=\xmax$. \label{fig:reset}}
\end{figure}

Figures \ref{fig:lowa_4}--\ref{fig:higha_4} demonstrate that how often the system completely resets is closely linked to the dynamics of how $X$ and $g(X_{\rm th}, t)$ evolve, and therefore observables such as waiting times and glitch sizes. The fraction of glitches that fully reset the system, i.e. result in $g(X_{\rm th}, t_{n}^+) = \phi(X_{\rm th})$ after a glitch at time $t_n$, is plotted as a function of $\alpha$ and $\xmax$ in Figure \ref{fig:reset}. The smallest value of $\alpha$ where over half the glitches reset the system shifts from 0.5 to 0.3 to 0.15 as $\xmax$ decreases from 0.99 to 0.95 to 0.8 respectively. This is intuitive, as lower $\alpha$ results in longer waiting times, all else being equal, and the stress is more likely to exceed $\xmax$ by the time a glitch is triggered.

For $\xmax \geq 1$, $X$ never exceeds $\xmax$, and the system never completely resets. That is, the repinning step in equation \eqref{eq:grepin} assigns vortices to pinning potentials with $X_{\rm th} \geq 1$, which never unpin. The system stagnates eventually, with all vortices pinned at sites whose stress thresholds cannot be reached. \citet{Melatos2009} ameliorated this stagnation by allowing a ``thermal creep'' term, whereby a small fraction of vortices at sites with $X_{\rm th} > X$ unpin randomly at a glitch \citep{Sneppen1997}. 

\subsection{Differences from previous meta-models}
The recipe for updating $g(X_{\rm th},t)$ and hence generating $\Delta X^{(n)}$ differs fundamentally from the recipe in previous meta-models \citep{Fulgenzi2017, Carlin2019quasi, Carlin2020}. Firstly $\Delta X^{(n)}$ is generated from $g(X_{\rm th},t)$, which is a new step. In the SDP and BSA meta-models, $\Delta X^{(n)}$ is drawn from an exogenously defined conditional jump distribution, $\eta[\Delta X^{(n)}\,|\,X(t_{n}^-)]$, which makes no reference to which pinning sites are occupied. Secondly, $\Delta X^{(n)}$ is generated deterministically; once $g(X_{\rm th},t_{n}^+)$ and $X(t_{n+1}^-)$ are known, so is $\Delta X^{(n)}$ without rolling dice. This fundamentally alters the meta-model from a doubly stochastic process, where the waiting time $\Delta t$ and avalanche size $\Delta X^{(n)}$ are drawn randomly from independent PDFs (inhomogeneous Poisson and $\eta[\Delta X^{(n)}\,|\,X(t_{n}^-)]$ respectively), to a singly stochastic process, where only $\Delta t$ is drawn randomly, according to the stress accumulation process of choice. 

With that said, the sequence of avalanche sizes is still unpredictable in a long-term sense in the endogenous-$\eta$ meta-model, because it is driven by stochastic draws of $\Delta t$. Furthermore, the ensemble average $g_s(X_{\rm th}) = \langle g(X_{\rm th}, t_n^+)\rangle$, is analogous to the statistically stationary global stress PDF $p(X)$ discussed in Section 6 of \citet{Fulgenzi2017}. From $g_s(X_{\rm th})$, it is possible to calculate an effective $\eta[\Delta X^{(n)}\,|\,X(t_{n}^-)]$, which is also statistically stationary. We do so in Appendix \ref{sec:gs}. If the goal of the theory is to predict the stationary glitch size and waiting time PDFs, then running the SDP meta-model as developed in previous papers and drawing randomly from $\eta[\Delta X^{(n)}\,|\,X(t_{n}^-)]$ derived from $g_s(X_{\rm th})$ as in Appendix \ref{sec:gs} is adequate. However, if the exact temporal sequence of sizes and waiting times is of interest, e.g. to investigate cross- and autocorrelations, then the evolution of $g(X_{\rm th}, t)$ must be tracked according to equation \eqref{eq:geom} or some variant thereof.

\subsection{Mutual friction} \label{sec:mutualfric}
In hydrodynamical multi-component models, the coupling between the inviscid and viscous components is called the mutual friction \citep{Hall1956, Barenghi1983, AnderssonSidery2006}. The strength and functional form of the mutual friction is fundamentally linked to how the vortices transfer angular momentum from the unobservable stress reservoir and the observable crust \citep{AnderssonSidery2006, Graber2018, Celora2020}. The SDP meta-model, and endogenous-$\eta$ extension considered in this paper, tacitly assume that mutual friction is weak between glitches as the stress grows linearly, irrespective of its absolute magnitude, until a glitch is triggered. These meta-models align qualitatively with phenomenological models that abruptly change the form (and usually weaken the strength) of the mutual friction when a glitch occurs, due to a phase transition between turbulent and laminar flow states \citep{Peralta2006, Mongiovi2017, Haskell2020}. We note that the meta-model treats glitches as impulsive events and has nothing to say in its present form about post-glitch recoveries, where hydrodynamic effects are likely to play a role.

\section{Observable long-term statistics} \label{sec:obs}
Following the steps outlined in Section \ref{sec:mcsteps} one can generate sequences of waiting times and sizes of arbitrary length, given $\alpha$, $K$, and $\xmax$. From these sequences, observable long-term statistics are predictable. They offer a baseline for falsification studies involving astronomical observations \citep{Melatos2018, Carlin2019quasi, Carlin2019ac}. As stated in equation \eqref{eq:kf}, we have $\Delta X = -KF \propto \Delta \nu$, implying the shape of the event size PDF $p(\Delta X)$ is the same as the observed glitch size PDF $p(\Delta \nu)$.

\subsection{Waiting time and size PDFs}

\begin{figure}
\centering
\includegraphics[width=0.9\linewidth]{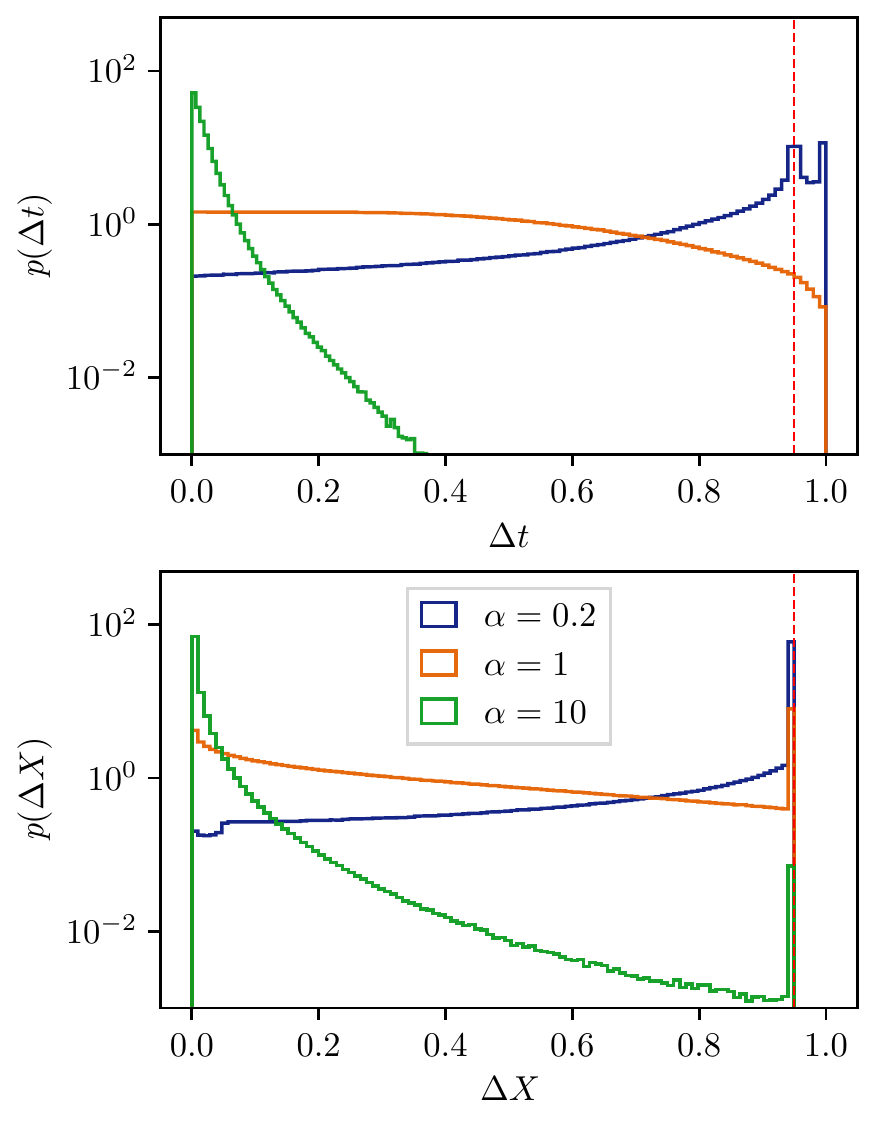}
\caption{Waiting time, $\Delta t$, and size, $\Delta X$, PDFs on the top and bottom panel respectively. Parameters: $N=10^7$ simulated glitches for each value of $\alpha$, $\xmax=0.95$ (indicated with a red dashed line in each panel), $K=\xmax$. \label{fig:pdfs_nolog}}
\end{figure}

Figure \ref{fig:pdfs_nolog} shows the waiting time PDF, $p(\Delta t)$, and size PDF, $p(\Delta X)$, for three values of $\alpha$ on a log-linear scale. For $\alpha \lesssim 0.5$, $p(\Delta t)$ has two peaks, one at $\Delta t = \xmax$ and one at $\Delta t = 1$, and increases monotonically for $\Delta t < \xmax$. In the low-$\alpha$ regime $X$ regularly exceeds $\xmax$, triggering a glitch of size $\Delta X = \xmax$. With $X \approx 1$ before the glitch, and $X \approx 1 - \xmax$ after, the maximum $\Delta t$ until the next glitch is $\xmax$. As equation \eqref{eq:pdelt} increases monotonically with $\Delta t$ for $\alpha < 1$, the maximum $\Delta t$ allowed by equation \eqref{eq:pdelt} is the most likely waiting time. When plotted on a log-log scale, we see that for $\alpha \gtrsim 3$, $p(\Delta t)$ is well approximated by a power-law distribution, with a turn-off at $\Delta t \lesssim 10^{-2}$ (in units of $\xc I_{\rm C} / N_{\rm em}$). Where this turn-off occurs depends on both $\alpha$ and $\xmax$. The slope of the power-law component of $p(\Delta t)$ is approximately $-3$ for $\alpha=10$ and decreases to $-4$ for $\alpha=50$. In the intermediate regime $0.5 \lesssim \alpha \lesssim 3$, $p(\Delta t)$ is approximately uniform.

The size PDFs also display distinctive features in the low- and high-$\alpha$ regimes. For $\alpha \lesssim 0.5$, $p(\Delta X)$ increases monotonically, with a significant fraction of the probability mass close to $\Delta X = \xmax$. The latter events correspond to the glitches that completely reset the system, if one has $X \geq \xmax$ prior to the glitch. On a log-log scale, we see that for $\alpha \gtrsim 3$, $p(\Delta X)$ can be approximated as a broken power law distribution, with a smooth turn-over at $\Delta X \approx 10^{-2}$ (in units of $\xc$). The location of this turn-over depends on $\alpha$ and $\xmax$. The slope of $p(\Delta X)$ for $\Delta X \gtrsim 10^{-2}$ is approximately $-2$ for $\alpha=10$, and decreases to $-2.5$ for $\alpha=50$. For $0.5 \lesssim \alpha \lesssim 3$, $p(\Delta X)$ is approximately uniform, with a spike at $\Delta X = \xmax$, again corresponding to glitches that completely reset the system.

In summary, in the low-$\alpha$ regime $p(\Delta t)$ is bimodal with peaks at $\Delta t = \xmax$ and $\Delta t = 1$, while $p(\Delta X)$ is monotonically increasing until $\Delta X = \xmax$ where there is a spike in the PDF then a sharp cut-off. In the high-$\alpha$ regime both $p(\Delta t)$ and $p(\Delta X)$ are approximately power-law distributed, with low-end turn-offs at $\Delta t \approx 10^{-2}$ and $\Delta X \approx 10^{-2}$.

\subsection{Cross- and autocorrelations}
\begin{figure}[t]
\centering
\includegraphics[width=0.82\linewidth]{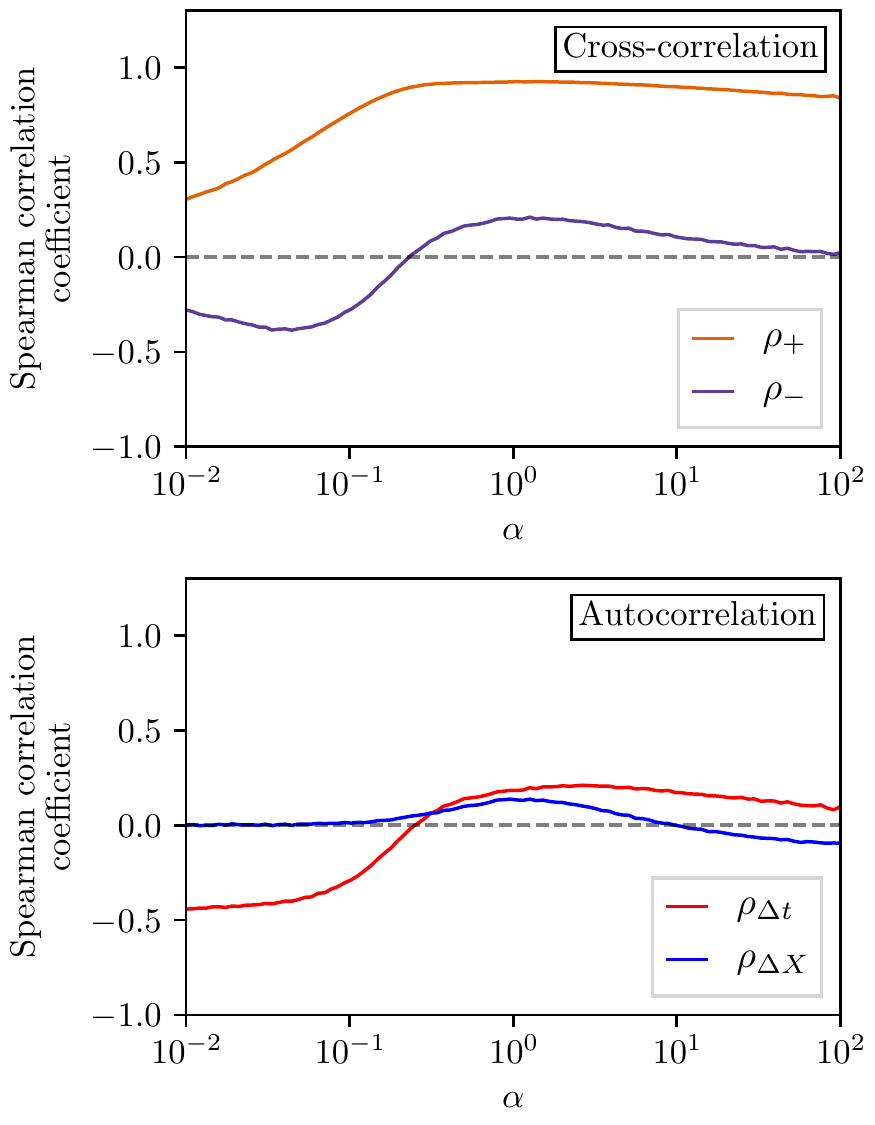}
\caption{Top panel: Forward cross-correlation, $\rho_+$, and backward cross-correlation, $\rho_-$, as a function of $\alpha$ (orange and purple curves respectively). Bottom panel: Auto-correlation between waiting times, $\rho_{\Delta t}$, and sizes, $\rho_{\Delta X}$, as a function of $\alpha$ (red and blue curves respectively). Parameters: $N=10^5$ glitches generated at 100 logarithmically spaced values of $\alpha$, $\xmax=0.95$, $K=\xmax$. \label{fig:all_correl}}
\end{figure}

The time-ordered nature of waiting times and relaxation events invites investigation into what relationships exist between events that happen consecutively. For example, in the SDP meta-model certain combinations of observables, such as the rate of spin-down and average waiting time, predict the cross-correlation between waiting times and the next glitch size \citep{Melatos2018}. In the SDP and BSA meta-models combinations of cross-correlations and autocorrelations restrict possible values of meta-model control parameters in individual glitching pulsars \citep{Carlin2019ac, Carlin2020}. 

Figure \ref{fig:all_correl} shows the Spearman correlation coefficients for the forward cross-correlation, $\rho_+$, i.e. the correlation between the size of the previous glitch and the next waiting time; the backward cross-correlation, $\rho_-$, i.e. the correlation between the size of the glitch and the preceding waiting time; as well as the autocorrelations between consecutive waiting times, $\rho_{\Delta t}$, and sizes, $\rho_{\Delta X}$. It is clear that $\rho_+$ is high for all but the smallest values of $\alpha$. The longer the waiting time, the larger the upper terminal in equation \eqref{eq:f}, and thus a greater fraction of the vortices unpin. At small values $\alpha \lesssim 10^{-2}$ most glitches completely reset the system, and the forward cross-correlation is lower. The backward cross-correlation, $\rho_-$, is small and negative for $\alpha \lesssim 0.2$, but small and positive for $\alpha \gtrsim 0.2$. Size autocorrelations, $\rho_{\Delta X}$, are negligible for all $\alpha$. Waiting time autocorrelations satisfy $\rho_{\Delta t} \sim -0.5$ for $\alpha \lesssim 0.1$, but are negligible for $\alpha \gtrsim 0.2$.

The impact of decreasing $\xmax$ on Figure \ref{fig:all_correl} is minimal; the magnitudes of the Spearman correlation coefficients decrease marginally, and the Spearman correlation coefficients as functions of $\alpha$ shift towards the right a small amount. As $\xmax$ decreases, Figure \ref{fig:reset} shows that more glitches reset the system, at a fixed value of $\alpha$.

\subsection{Aftershocks and precursors} \label{sec:aftershock}
\begin{figure}[t]
\centering
\includegraphics[width=0.82\linewidth]{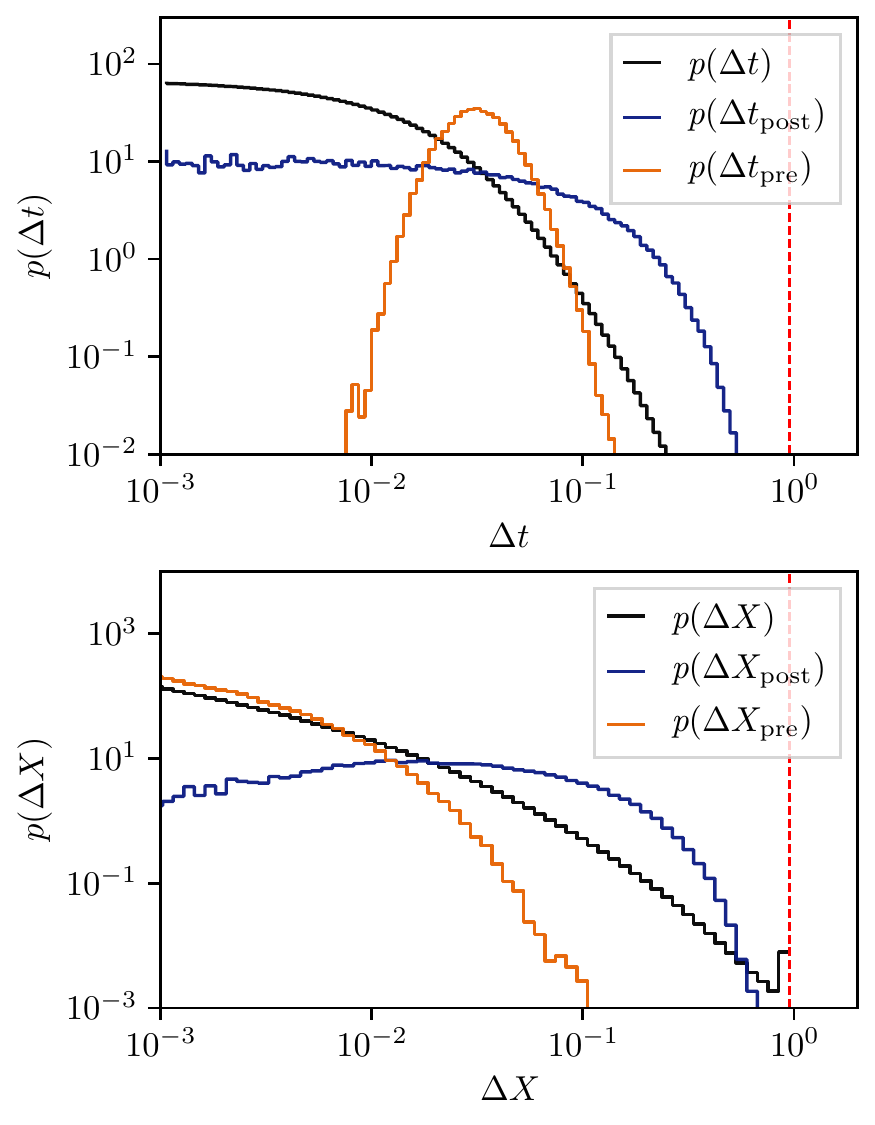}
\caption{Top panel: PDFs of all waiting times, $\Delta t$, waiting times just prior to a system-resetting glitch, $\Delta t_{\rm pre}$, and waiting times just after a system-resetting glitch, $\Delta t_{\rm post}$. Bottom panel: Corresponding PDFs for sizes. Parameters: $\alpha=10$, $N=10^8$ total glitches, $\sim10^5$ of which reset the system completely, $\xmax=0.95$, $K=\xmax$. \label{fig:aftershocks}}
\end{figure}

Aftershocks occur after a large event, when subsequent events are larger and more frequent than usual. They are common in spatially correlated knock-on processes such as avalanches and earthquakes \citep{Kagan1991, Utsu1995, Jensen1998}. Naively, one might expect that the endogenous-$\eta$ meta-model should not exhibit aftershocks, as it does not include spatial correlations, due to the assumption that each pinning site experiences the same spatially-averaged stress. Aftershocks are not discussed in the context of the SDP or BSA meta-models, as large events in those meta-models do not correspond to a rearrangement of occupied pinning sites. A large event does not impact the next waiting time or size, beyond reducing the stress in the system.

We investigate whether the above naive expectation holds for the endogenous-$\eta$ meta-model by calculating the conditional waiting time and size PDFs for glitches immediately following a glitch that completely resets the system, $p(\Delta t_{\rm post})$ and $p(\Delta X_{\rm post})$ respectively. Figure \ref{fig:aftershocks} shows these PDFs, and the waiting time and size PDFs of events just prior to a glitch that completely resets the system, $p(\Delta t_{\rm pre})$ and $p(\Delta X_{\rm pre})$ respectively. There is an excess of longer waiting times in $p(\Delta t_{\rm post})$ compared to $p(\Delta t)$, as the system takes some time to build up stress before another glitch is likely, after such a system-resetting glitch. There is a corresponding excess of larger sizes in $p(\Delta X_{\rm post})$ compared to $p(\Delta X)$. The precursor waiting times, $\Delta t_{\rm pre}$, are distributed as a log-normal distribution, with $\langle\Delta t_{\rm pre}\rangle > \langle \Delta t \rangle$. There are relatively few small $\Delta t_{\rm pre}$ events, because the system-resetting condition $\Delta X \geq \xmax$ occurs only if enough stress accumulates before the glitch. There are few large glitches in $p(\Delta X_{\rm pre})$ compared to $p(\Delta X)$, as large glitches place the system in a configuration which is unlikely to fully reset at the next glitch, as $X$ is low. 

The behavior of aftershocks and precursors in Figure \ref{fig:aftershocks} is replicated qualitatively for other values of $\alpha \neq 10$ and $\xmax \neq 0.95$. That is, there is always an excess of longer waiting times, and larger glitches, following a system-resetting glitch, while the precursor waiting times are longer than average. For $\alpha \lesssim 0.2$ these features in the PDFs are less prominent, as the system resets after almost every event.

\section{Astronomical observations} \label{sec:disc}
How do the long-term statistical predictions in Section \ref{sec:obs} compare to what we currently see in glitching pulsars? This question has been answered previously in the context of the SDP \citep{Melatos2018, Carlin2019quasi, Carlin2019ac}, and BSA \citep{Carlin2020} meta-models. For the five pulsars with the most recorded glitches, there are regimes of parameter space in the SDP meta-model which adequately explain observations. For example, PSR J0537$-$6910 is consistent with the SDP meta-model if $\alpha \lesssim 0.1$ \citep{Carlin2019ac}. However this is not the case for the BSA meta-model, for which two pulsars, namely PSR J0534$+$2200 and PSR J1341$-$6220, are consistent with the meta-model only if there exists an undetected population of frequent small glitches. 

The endogenous-$\eta$ meta-model presented here is falsifiable using the same approach: the meta-model must predict simultaneously, with one set of input parameters, the long-term $p(\Delta t)$, $p(\Delta X)$, cross-correlations, and autocorrelations, otherwise it is rejected. In principle we have one additional potential observable in this meta-model, the aftershocks and precursors discussed in Section \ref{sec:aftershock}. However, due to the small number of glitches recorded in individual pulsars ($N < 50$), it is not yet clear whether we have witnessed any large, system-resetting glitches in astronomical data, and so we cannot compare to the aftershock or precursor predictions of Section \ref{sec:aftershock}. 

Current observations of glitch waiting time and size distributions \citep{Howitt2018, Fuentes2019} do not show clear signs of a sharp cut-off at the upper end, as the meta-model predicts in Figure \ref{fig:pdfs_nolog} for $\alpha \lesssim 5$. For $\alpha \gtrsim 5$, where the distributions are steeper, the sharp cut-off may have escaped detection until now due to the paucity of recorded glitches. Even so, if most glitching pulsars fall in the $\alpha \gtrsim 5$ regime we should expect to see significant forward cross-correlations, $\rho_+ \gtrsim 0.8$, in more pulsars, as seen in Figure \ref{fig:all_correl}. Yet only two pulsars, PSR J0537$-$6910 and PSR J1341$-$6220 have $\rho_+$ significantly different from zero, at 95\% confidence. Neither of these objects favor a power-law glitch size distribution, as the model predicts for $\alpha \gtrsim 5$. We therefore conclude provisionally that the endogenous-$\eta$ meta-model is incompatible with existing pulsar observations, although more data are needed to be confident of course. This is an important result, because the endogenous-$\eta$ meta-model codifies the traditional, popular understanding in the literature regarding how vortex pinning and unpinning occurs stochastically, as explained in Sections \ref{sec:intro}--\ref{sec:sdp}.

Model parameters such as $\alpha$, the dimensionless control parameter that determines the speed at which stress accumulates, and $K$, the coupling constant between changes in stress and the observable change in frequency, may be partially informed by independent (i.e. non-glitch) observations of pulsars. As discussed in Section 3 of \citet{Melatos2018}, and Section 5 of \citet{Carlin2020}, we have \begin{eqnarray}
\alpha \approx \frac{\xc}{\langle\Delta t \rangle \dot{\nu}}\ , \label{eq:alpha}
\end{eqnarray}
up to a factor of order unity, where $\langle\Delta t \rangle$ is the mean waiting time between glitches, and $\dot{\nu}$ is the long-term spin-down rate of the pulsar, after correcting for glitches and timing noise. The critical stress at which a glitch becomes certain, $\xc$, depends on the equation of state and is unknown in general \citep{Link1991, Donati2006}. Nevertheless, the denominator in Equation \eqref{eq:alpha} varies between pulsars by many orders of magnitude; see Table 2 of \citet{Melatos2018}, for example. The coupling constant $K$ has even more uncertainties surrounding it, as described in Section \ref{sec:unpin}. If a linear coupling between the crust and the stress is assumed at a glitch, we recover Equation \eqref{eq:kr}, however non-linear couplings are also plausible, and may easily change the dependence of $K$ on observables \citep{Gugercinoglu2019, Pizzochero2020, Celora2020}. A preliminary exploration of how $K$ impacts the endogenous-$\eta$ meta-model is presented in Appendix \ref{sec:lowk}. 

Direct comparisons of the meta-model predictions to long-term statistics derived from glitch catalogues are predicated on the assumption that all glitches are detected, i.e. that the datasets are complete. \citet{Espinoza2014} and \citet{Espinoza2020} claimed that this assumption is true for PSR J0534$+$2200 and PSR0835$-$4510 respectively. Monte Carlo injection studies have been published which seek to quantify completeness, but they are hampered by human intervention in traditional glitch finding algorithms \citep{Janssen2006, Yu2017b}. Autonomous glitch finding algorithms, such as those based on hidden Markov models \citep{Melatos2020hmm} and nested sampling \citep{Shannon2016, Jankowski2019, Lower2020}, will probe this question quantitatively for more pulsars, especially as more pulsars are regularly timed.

\newpage
\section{Conclusion} \label{sec:concl}
The long-term predictions from meta-models of stress-relax processes are avenues to falsify otherwise plausible microphysical mechanisms that the meta-model encompasses. The SDP meta-model \citep{Fulgenzi2017} encompasses mechanisms wherein glitches are triggered probabilistically, and are more likely to occur when the system-wide stress is higher. The falsifiable predictions such as size and waiting-time cross-correlations \citep{Melatos2018}, and autocorrelations \citep{Carlin2019ac}, depend on the functional form of the conditional avalanche size PDF, $\eta[\Delta X^{(n)}\,|\,X(t_{n}^-)]$ \citep{Carlin2019quasi}. Previously, this PDF was fixed exogenously to be a power law \citep{Fulgenzi2017, Melatos2018, Melatos2019}, motivated by Gross-Pitaevskii simulations of vortex avalanches \citep{Warszawski2011}, and the sizes of stress-release avalanches in other self-organized critical systems \citep{Jensen1998}. However, when $\eta[\Delta X^{(n)}\,|\,X(t_{n}^-)]$ is a power law, the SDP meta-model does not explain the quasiperiodic waiting times and unimodal size PDF seen in some pulsars \citep{Howitt2018}. This is ameliorated by adjusting ``by hand'' $\eta[\Delta X^{(n)}\,|\,X(t_{n}^-)]$ to a unimodal distribution, whereupon the SDP meta-model remains consistent with the data \citep{Carlin2019quasi}.  

In this paper, we generalize the SDP meta-model so that $\eta[\Delta X^{(n)}\,|\,X(t_{n}^-)]$ is generated endogenously via the coherent stress mechanism \citep{Sneppen1997, Melatos2009}. The coherent stress mechanism encapsulates the traditional understanding of how superfluid vortex pinning and unpinning proceeds stochastically in a neutron star. The memory of previous glitches is imprinted on the PDF of occupied pinning sites, $g(X_{\rm th}, t)$, which we emphasize does not equal the distribution of available sites $\phi(X_{\rm th})$ in general. Vortices pinned at sites with thresholds below the stress at a glitch unpin when a glitch is triggered, repinning at sites with thresholds above the stress after the glitch. The size of the glitch is proportional to the fraction of vortices unpinned in this manner. 

The endogenous-$\eta$ meta-model produces a broad phenomenology of observable waiting time and size PDFs, conditional on the choice of control parameters. In all circumstance, though, it produces high forward cross-correlations between glitch sizes and subsequent waiting times ($\rho_+ \gtrsim 0.8$), which are largely absent from observational data. It also predicts that either \begin{enumerate*}[(i)]
\item the system stagnates, for $\xmax \geq \xc$, or 
\item there is an excess of the largest waiting times and sizes, corresponding to events which completely reset the system by unpinning all vortices. 
\end{enumerate*}
Associated with these system-resetting events are aftershocks, which are larger and occur later than an average glitch, and precursors, which are smaller than an average glitch. There is no evidence for such large, system-resetting events in the size or waiting time PDFs of any glitching pulsar. We therefore conclude provisionally that the endogenous-$\eta$ version of the SDP meta-model is falsified by existing data, although more data are needed to be sure.

The provisional falsification of the endogenous-$\eta$ version of the SDP meta-model is important. The coherent stress process is not just any process; it embodies the traditional understanding of how vortex pinning and unpinning works throughout the literature \citep{Haskell2015}. Moreover it is intriguing that the version of the SDP meta-model in which $\eta[\Delta X^{(n)}\,|\,X(t_{n}^-)]$ is specified exogenously is not falsified by existing pulsar data. What are we to make of this situation? Is it a hint that some microphysics other than superfluid vortex avalanches is at work? Is the success (at escaping falsification) of the SDP meta-model with exogenous $\eta[\Delta X^{(n)}\,|\,X(t_{n}^-)]$ thanks simply to the flexibility it affords, when choosing $\eta[\Delta X^{(n)}\,|\,X(t_{n}^-)]$ selectively to suit every individual pulsar? It is too early to say. The endogenous-$\eta$ meta-model is still an idealized, phenomenological representation of what happens inside a pulsar during a superfluid vortex avalanche. For example, recent $N$-body point-vortex simulations of collective, glitch-like vortex motion indicates that the stress is spatially correlated \citep{Howitt2020}. Spatial correlations are notoriously hard to treat theoretically, but it is known that they can alter the observable statistics of a far-from-equilibrium system comprehensively, e.g. in self-organized critical systems \citep{Jensen1998, Aschwanden2018}. Larger and more complete glitch catalogs generated by the latest generation of glitch monitoring campaigns at radio wavelengths are likely to play a central role in resolving some of the physical puzzles above \citep{Stappers2011, Jankowski2019}.

\section*{Acknowledgements}
Parts of this research are supported by the Australian Research Council (ARC) Centre of Excellence for Gravitational Wave Discovery (OzGrav) (project number CE170100004) and ARC Discovery Project DP170103625. JBC is supported by an Australian Postgraduate Award.

\newpage
\appendix
\section{Non-maximal stress-crust coupling} \label{sec:lowk}
\begin{figure}[t]
\centering
\includegraphics[width=0.7\linewidth]{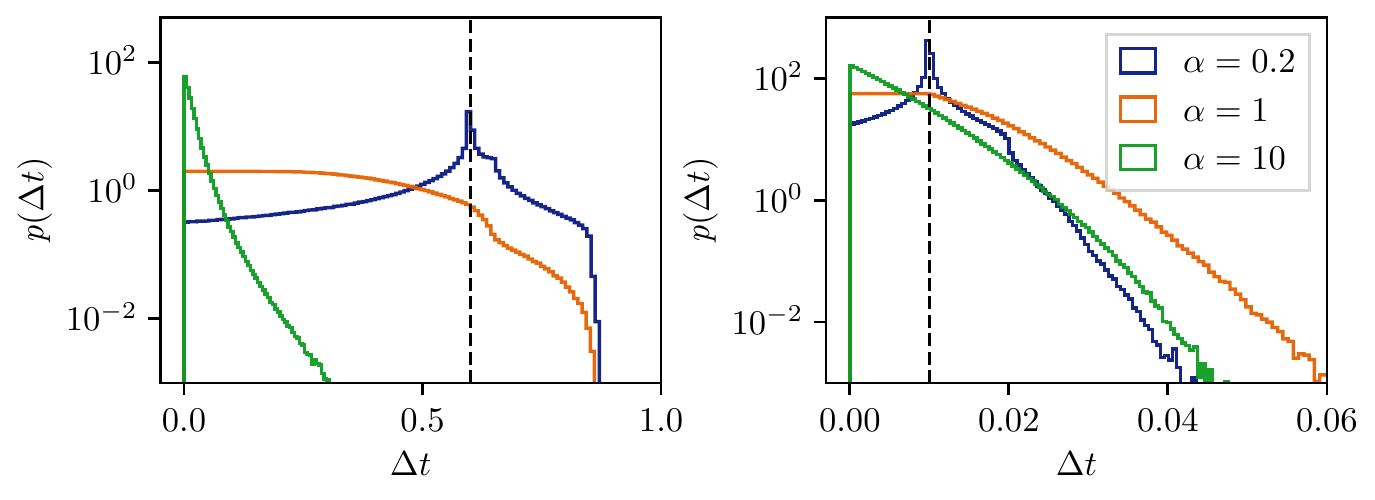}
\caption{Waiting time PDFs for $K=0.6$ (left panel) and $K=0.01$ (right panel). Parameters: $N=10^7$ simulated glitches for each value of $\alpha$, with $\xmax = 0.95$. The black dashed vertical line indicates $\Delta t = K$ in each panel. \label{fig:lowk}}
\end{figure}

The coupling factor $K$ defined in equation \eqref{eq:delx} and appearing in equations \eqref{eq:kf} and \eqref{eq:kr} relates changes in the stress variable $X$ to the angular velocity $2\pi\nu$ of the crust. The long-term observables in Section 4 are predicted assuming $K$ takes its maximum value, $\xmax$. Somewhat counter-intuitively, the long-term statistics do change --- modestly, but meaningfully --- when non-maximal coupling is considered.

One always has $K \leq \xmax$ to ensure $X(t)\geq 0$. In Section \ref{sec:unpin}, we specialize to $K= \xmax$ for the sake of definiteness. In this appendix we briefly check the case $K < \xmax$. Figure \ref{fig:lowk} shows the PDF of waiting times, $p(\Delta t)$, for two representative values of $K < \xmax = 0.95$. For $K=0.6$, $p(\Delta t)$ looks broadly similar to the top panel of Figure \ref{fig:pdfs_nolog} ($K=\xmax$), except for $\alpha=0.2$, which no longer has a peak at $\Delta t = 1$. Long waiting times are suppressed as $K$ sets the maximum event size in equation \eqref{eq:kf}. Smaller $K$ increases the minimum stress in the system, as less stress is released at the largest glitches (i.e. when $F=1$). For $K=0.01$ and $\alpha\gtrsim1$, $p(\Delta t)$ is exponentially distributed. For $\alpha\lesssim1$, $p(\Delta t)$ is peaked around $\Delta t = K$, but has an exponential tail. The size PDFs, $p(\Delta X)$, are broadly the same shape as in the bottom panel of Figure \ref{fig:pdfs_nolog} ($K = \xmax$), but for $K < \xmax$ and $\alpha \lesssim 1$ they are more strongly peaked around $\Delta X = K$, and for $\alpha \gtrsim 1$ they are power-law distributed over many decades, with a small peak at $\Delta X = K$. In summary, if $K < \xmax$, in the high-$\alpha$ regime $p(\Delta t)$ is exponentially distributed and $p(\Delta X)$ is power-law distributed, while in the low-$\alpha$ regime $p(\Delta t)$ is unimodal around a peak at $\Delta t = K$ and $p(\Delta X)$ is sharply peaked at $\Delta X = K$.

The behavior of the cross-correlations and autocorrelations as a function of $\alpha$ does not change for $K < \xmax$ compared to $K = \xmax$, beyond shifting the features seen in Figure \ref{fig:all_correl} to the right, e.g. for $K=0.1$ the peak in $\rho_+$ occurs at $\alpha \approx 10$ instead of $\alpha \approx 0.5$. Reducing $K$ has a similar impact as reducing $\xmax$ on the fraction of events that completely reset the system, as shown in Figure \ref{fig:reset}. At a fixed $\alpha$, reducing $K$ reduces the amount of stress released. Hence we reach $X \geq \xmax$ more often by the time the next glitch is triggered. 

\section{Ensemble-averaged stress threshold distribution at occupied pinning sites} \label{sec:gs}
The PDF $g(X_{\rm th}, t)$ is a stochastically fluctuating function, dependent on the random sequence of waiting times drawn up to time $t$, as described in Section \ref{sec:eom}. A stationary analogue, $g_s(X_{\rm th})$, would allow for a priori prediction of long-term statistics that do not depend on the exact time-ordered nature of events, such as $p(\Delta t)$ and $p(\Delta X)$ \citep{Sneppen1997, Melatos2009}.

\begin{figure}[t]
\centering
\includegraphics[width=0.4\linewidth]{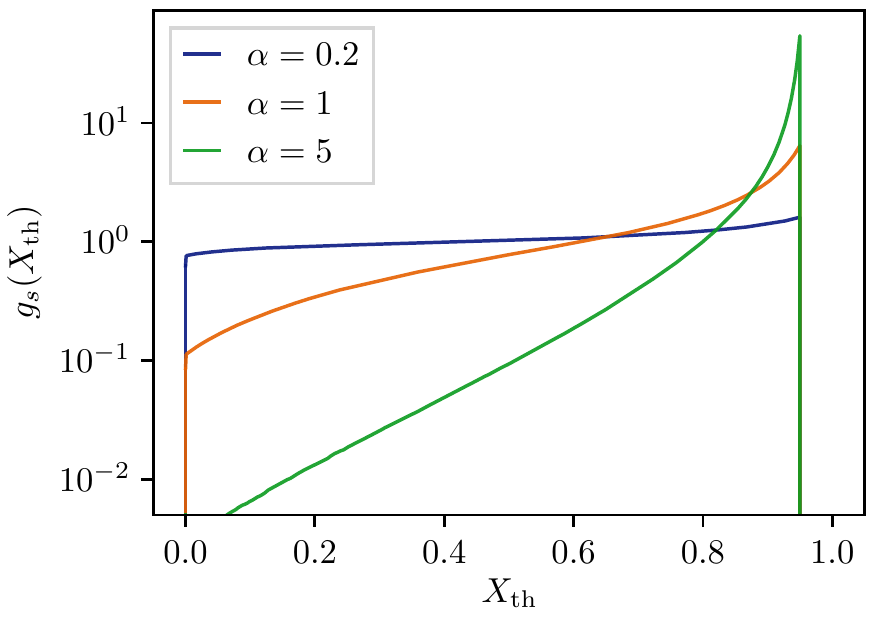}
\caption{Ensemble averaged PDF of occupied pinning sites $g_s(X_{\rm th}) = \langle g(X_{\rm th}, t) \rangle$ calculated empirically for three values of $\alpha$. Parameters: $10^7$ samples of $g(X_{\rm th})$ for each value of $\alpha$, $\xmax=0.95$, with $K=\xmax$. \label{fig:gs}}
\end{figure}

Following the notation in \citet{Daly2007} and \citet{Fulgenzi2017}, $p(\Delta t)$ and $p(\Delta X)$ are related to the PDFs $p_e(Y)\td Y$ and $\,p_s(X)\td X$, the probabilities that the stress is in $(Y, Y+\td Y)$ just before a glitch and in $(X, X+\td X)$ just after a glitch respectively. With these definitions we obtain
\begin{eqnarray}
p(\Delta t) &= \int_0^{1 - \Delta t} \td Y p_s(Y) \lambda(Y + \Delta t) \exp[-\Lambda(Y + \Delta t) \Lambda(Y)]\ , \label{eq:pdelt_app}
\end{eqnarray}
and 
\begin{eqnarray}
p(\Delta X) = \int_{\Delta X}^{1} \td Y p_e(Y) \eta(\Delta X\,|\,Y)\ , \label{eq:pdelx_app}
\end{eqnarray}
with $\Lambda(x) = \int_0^x\td x' \lambda(x')$, and $\Delta X = Y - X$. In the endogenous-$\eta$ meta-model $\eta(\Delta X\,|\,Y)$, is a deterministic function of the occupied pinning site PDF, $g(X_{\rm th})$, as discussed in Section \ref{sec:unpin}. The relationship becomes probabilistic again, as in the SDP meta-model, if we consider the statistically stationary PDF, $g_s(X_{\rm th}) = \langle g(X_{\rm th}, t) \rangle$, where the brackets indicate an ensemble average. The PDF $g_s(X_{\rm th})$ is related to $\eta(\Delta X\,|\,Y)$ via 
\begin{eqnarray}
\eta(\Delta X\,|\,Y) = K \int_0^{p_e(Y)} \td X_{\rm th}\, g_s(X_{\rm th})\ . \label{eq:etags}
\end{eqnarray}
Solving equations \eqref{eq:pdelt_app}--\eqref{eq:etags} analytically for $g_s(X_{\rm th})$ is outside the scope of this paper. Instead, we show an estimate of $g_s(X_{\rm th})$ at three values of $\alpha$ in Figure \ref{fig:gs}, calculated by running the automaton outlined in Section \ref{sec:mcsteps} and sampling $g(X_{\rm th}, t)$ after every glitch. As expected, $g_s(X_{\rm th})$ is a strictly increasing function of $X_{\rm th}$, i.e. sites with a higher threshold are occupied more often than those with a lower threshold. For $\alpha \lesssim 1$, $g_s(X_{\rm th})$ is well approximated by a power law with upper and lower cut-offs, i.e. $g_s(X_{\rm th}) \propto X_{\rm th}^{\gamma}H(X_{\rm th})H(\xmax - X_{\rm th})$, with $0 < \gamma \lesssim 3$, and $\gamma$ growing with $\alpha$. 

\section{Relation to point processes and time series modeling} \label{sec:timeseries}
The endogenous-$\eta$ meta-model in this paper and its exogenous-$\eta$ alternatives, such as the SDP \citep{Fulgenzi2017} and BSA \citep{Carlin2020} meta-models, are examples of stochastic time series with jumps. We observe the jumps as a point process. In this appendix, we situate the endogenous-$\eta$ meta-model within canonical classification schemes for such processes. The classification schemes are not unique.

The SDP meta-model is an example of a doubly stochastic, marked point process \citep{Cox1966, Cinlar1975, Fulgenzi2017}. It is a one-dimensional point process because it is a sequence of instantaneous events ordered in time. Marked refers to additional information (i.e. the event size) that is associated with the epoch of each event. It is doubly stochastic as both the event sizes and the waiting times between events are random processes. The endogenous-$\eta$ meta-model is not doubly stochastic, as the event sizes are deterministic, after the waiting time is chosen (given a certain history of past avalanches), as described in Section \ref{sec:sec2}. For both the SDP and endogenous-$\eta$ meta-models, the state of the system jumps discontinuously at each event making them examples of a jump process. The SDP meta-model is Markovian, the current state only depends on the immediately previous state. However the endogenous-$\eta$ meta-model is not, as the threshold distribution of occupied pinning sites contains a long-lasting memory of states before the one immediately previous.

The analysis of point processes from a statistical perspective is often presented alongside time series modeling \citep{Kingman1993, Box2015}. Autoregressive models, such as the autoregressive integrated moving average (ARIMA) model, can be used to model a wide variety of astronomical time series data \citep{Feigelson2018}. Adopting the formalism of autoregressive models unlocks a large, well-tested literature for common tasks such as maximum likelihood parameter estimation, model comparison, and model validation \citep{Box2015}. Recent advances in filters for hidden semi-Markov models are another intriguing avenue, as these models are explicitly designed to track a process that jumps between states at irregularly spaced intervals \citep{VanDerHoek2018}.

In this paper and others investigating glitch meta-models, we elect not to analyze the problem in terms of autoregressive models for three reasons. \begin{enumerate*}[(i)] \item Glitch sample sizes are small ($N \lesssim 50$), so there is no practical imperative to exploit the computational efficiency offered by autoregressive models. 
\item We are motivated by the astrophysical goal of exploring the long-term statistical behavior of an automaton which formalizes directly the popular intuitive picture of glitches as a stick-slip, stress-relax phenomenon. It is possible in principle to formulate the stress-relax dynamics less directly in the language of an autoregressive model, and we leave that for future work.
\item Some work has been done on parameter estimation by Markov chain Monte Carlo methods for the SDP meta-model \citep{Melatos2019}. Extending this work to other meta-models or larger data sets may benefit from the computational efficiency of autoregressive modeling, but is outside the scope of this work. 
\end{enumerate*}

\bibliographystyle{aasjournal}
\bibliography{endog_eta_bib}

\begin{thebibliography}{}
\expandafter\ifx\csname natexlab\endcsname\relax\def\natexlab#1{#1}\fi
\providecommand{\url}[1]{\href{#1}{#1}}
\providecommand{\dodoi}[1]{doi:~\href{http://doi.org/#1}{\nolinkurl{#1}}}
\providecommand{\doeprint}[1]{\href{http://ascl.net/#1}{\nolinkurl{http://ascl.net/#1}}}
\providecommand{\doarXiv}[1]{\href{https://arxiv.org/abs/#1}{\nolinkurl{https://arxiv.org/abs/#1}}}

\bibitem[{Alpar {et~al.}(1996)}]{Alpar1996}
Alpar, M.~A., {et~al.} 1996, Astrophysical Journal, 459, 706,
  \dodoi{10.1086/176935}

\bibitem[{Anderson \& Itoh(1975)}]{Anderson1975}
Anderson, P.~W., \& Itoh, N. 1975, Nature, 256, 25, \dodoi{10.1038/256025a0}

\bibitem[{Andersson {et~al.}(2003)Andersson, Comer, \& Prix}]{Andersson2003}
Andersson, N., Comer, G.~L., \& Prix, R. 2003, Phys. Rev. Lett., 90, 091101,
  \dodoi{10.1103/PhysRevLett.90.091101}

\bibitem[{Andersson {et~al.}(2006)Andersson, Sidery, \&
  Comer}]{AnderssonSidery2006}
Andersson, N., Sidery, T., \& Comer, G.~L. 2006, Monthly Notices of the Royal
  Astronomical Society, 368, 162, \dodoi{10.1111/j.1365-2966.2006.10147.x}

\bibitem[{Andersson {et~al.}(2012)}]{Andersson2012}
Andersson, N., {et~al.} 2012, Phys. Rev. Lett., 109, 241103,
  \dodoi{10.1103/PhysRevLett.109.241103}

\bibitem[{Aschwanden {et~al.}(2018)}]{Aschwanden2018}
Aschwanden, M.~J., {et~al.} 2018, Space Sci Rev, 214, 55,
  \dodoi{10.1007/s11214-018-0489-2}

\bibitem[{Ashton {et~al.}(2019)}]{Ashton2019}
Ashton, G., {et~al.} 2019, Nat Astron, 1, \dodoi{10.1038/s41550-019-0844-6}

\bibitem[{Avogadro {et~al.}(2007)}]{Avogadro2007}
Avogadro, P., {et~al.} 2007, Physical Review C, 75, 012805,
  \dodoi{10.1103/PhysRevC.75.012805}

\bibitem[{Barenghi {et~al.}(1983)Barenghi, Donnelly, \& Vinen}]{Barenghi1983}
Barenghi, C.~F., Donnelly, R.~J., \& Vinen, W.~F. 1983, Journal of Low
  Temperature Physics, 52, 189, \dodoi{10.1007/BF00682247}

\bibitem[{Box {et~al.}(2015)}]{Box2015}
Box, G. E.~P., {et~al.} 2015, Time {{Series Analysis}}: {{Forecasting}} and
  {{Control}} ({Hoboken, New Jersey}: {John Wiley and Sons Inc.})

\bibitem[{Carlin \& Melatos(2019{\natexlab{a}})}]{Carlin2019quasi}
Carlin, J.~B., \& Melatos, A. 2019{\natexlab{a}}, MNRAS, 483, 4742,
  \dodoi{10.1093/mnras/sty3433}

\bibitem[{Carlin \& Melatos(2019{\natexlab{b}})}]{Carlin2019ac}
---. 2019{\natexlab{b}}, MNRAS, 488, 4890, \dodoi{10.1093/mnras/stz2014}

\bibitem[{Carlin \& Melatos(2020)}]{Carlin2020}
---. 2020, Mon Not R Astron Soc, 494, 3383, \dodoi{10.1093/mnras/staa935}

\bibitem[{Celora {et~al.}(2020)}]{Celora2020}
Celora, T., {et~al.} 2020, Monthly Notices of the Royal Astronomical Society,
  496, 5564, \dodoi{10.1093/mnras/staa1930}

\bibitem[{Chamel \& Haensel(2008)}]{Chamel2008}
Chamel, N., \& Haensel, P. 2008, Living Rev. Relativ., 11, 10,
  \dodoi{10.12942/lrr-2008-10}

\bibitem[{Chugunov \& Horowitz(2010)}]{Chugunov2010}
Chugunov, A.~I., \& Horowitz, C.~J. 2010, Monthly Notices of the Royal
  Astronomical Society: Letters, 407, 54,
  \dodoi{10.1111/j.1745-3933.2010.00903.x}

\bibitem[{{\c C}inlar(1975)}]{Cinlar1975}
{\c C}inlar, E. 1975, Introduction to Stochastic Processes ({Englewood Cliffs,
  N.J. : Prentice-Hall})

\bibitem[{Cox(1955)}]{Cox1955}
Cox, D.~R. 1955, Journal of the Royal Statistical Society. Series B, 17, 129,
  \dodoi{10.2307/2983950}

\bibitem[{Cox \& Lewis(1966)}]{Cox1966}
Cox, D.~R., \& Lewis, P. A.~W. 1966, The {{Statistical Analysis}} of {{Series}}
  of {{Events}} ({Chapman and Hall})

\bibitem[{Daly \& Porporato(2007)}]{Daly2007}
Daly, E., \& Porporato, A. 2007, Physical Review E, 75, 11119,
  \dodoi{10.1103/PhysRevE.75.011119}

\bibitem[{Dodson {et~al.}(2002)Dodson, McCulloch, \& Lewis}]{Dodson2002}
Dodson, R.~G., McCulloch, P.~M., \& Lewis, D.~R. 2002, The Astrophysical
  Journal Letters, 564, L85, \dodoi{10.1086/339068}

\bibitem[{Donati \& Pizzochero(2006)}]{Donati2006}
Donati, P., \& Pizzochero, P.~M. 2006, Physics Letters B, 640, 74,
  \dodoi{10.1016/j.physletb.2006.07.047}

\bibitem[{Espinoza {et~al.}(2011)}]{Espinoza2011}
Espinoza, C.~M., {et~al.} 2011, Monthly Notices of the Royal Astronomical
  Society, 414, 1679, \dodoi{10.1111/j.1365-2966.2011.18503.x}

\bibitem[{Espinoza {et~al.}(2014)}]{Espinoza2014}
---. 2014, Monthly Notices of the Royal Astronomical Society, 440, 2755,
  \dodoi{10.1093/mnras/stu395}

\bibitem[{Espinoza {et~al.}(2020)}]{Espinoza2020}
---. 2020, arXiv:2007.02921 [astro-ph].
\newblock \doarXiv{2007.02921}

\bibitem[{Feigelson {et~al.}(2018)Feigelson, Babu, \& Caceres}]{Feigelson2018}
Feigelson, E.~D., Babu, G.~J., \& Caceres, G.~A. 2018, Frontiers in Physics, 6,
  80, \dodoi{10.3389/fphy.2018.00080}

\bibitem[{Field {et~al.}(1995)}]{Field1995}
Field, S., {et~al.} 1995, Physical Review Letters, 74, 1206,
  \dodoi{10.1103/PhysRevLett.74.1206}

\bibitem[{Fuentes {et~al.}(2019)Fuentes, Espinoza, \&
  Reisenegger}]{Fuentes2019}
Fuentes, J.~R., Espinoza, C.~M., \& Reisenegger, A. 2019, A\&A, 630, A115,
  \dodoi{10.1051/0004-6361/201935939}

\bibitem[{Fulgenzi {et~al.}(2017)Fulgenzi, Melatos, \& Hughes}]{Fulgenzi2017}
Fulgenzi, W., Melatos, A., \& Hughes, B.~D. 2017, MNRAS, 470, 4307,
  \dodoi{10.1093/mnras/stx1353}

\bibitem[{Glampedakis \& Andersson(2009)}]{Glampedakis2009}
Glampedakis, K., \& Andersson, N. 2009, Physical Review Letters, 102, 141101,
  \dodoi{10.1103/PhysRevLett.102.141101}

\bibitem[{Graber {et~al.}(2018)Graber, Cumming, \& Andersson}]{Graber2018}
Graber, V., Cumming, A., \& Andersson, N. 2018, ApJ, 865, 23,
  \dodoi{10.3847/1538-4357/aad776}

\bibitem[{G{\"u}gercino{\u g}lu \& Alpar(2019)}]{Gugercinoglu2019}
G{\"u}gercino{\u g}lu, E., \& Alpar, M.~A. 2019, Monthly Notices of the Royal
  Astronomical Society, 488, 2275, \dodoi{10.1093/mnras/stz1831}

\bibitem[{Hall \& Vinen(1956)}]{Hall1956}
Hall, H.~E., \& Vinen, W.~F. 1956, Proceedings of the Royal Society of London
  Series A, 238, 215, \dodoi{10.1098/rspa.1956.0215}

\bibitem[{Haskell(2016)}]{Haskell2016}
Haskell, B. 2016, Monthly Notices of the Royal Astronomical Society: Letters,
  461, L77, \dodoi{10.1093/mnrasl/slw103}

\bibitem[{Haskell \& Antonopoulou(2014)}]{Haskell2014}
Haskell, B., \& Antonopoulou, D. 2014, Monthly Notices of the Royal
  Astronomical Society, 438, L16, \dodoi{10.1093/mnrasl/slt146}

\bibitem[{Haskell {et~al.}(2020)Haskell, Antonopoulou, \&
  Barenghi}]{Haskell2020}
Haskell, B., Antonopoulou, D., \& Barenghi, C. 2020, Monthly Notices of the
  Royal Astronomical Society, 499, 161, \dodoi{10.1093/mnras/staa2678}

\bibitem[{Haskell \& Melatos(2015)}]{Haskell2015}
Haskell, B., \& Melatos, A. 2015, International Journal of Modern Physics D,
  24, 1530008, \dodoi{10.1142/S0218271815300086}

\bibitem[{Haskell \& Melatos(2016)}]{Haskell2016a}
---. 2016, Monthly Notices of the Royal Astronomical Society, 461, 2200,
  \dodoi{10.1093/mnras/stw1334}

\bibitem[{Hooker {et~al.}(2015)Hooker, Newton, \& Li}]{Hooker2015}
Hooker, J., Newton, W.~G., \& Li, B.-A. 2015, Monthly Notices of the Royal
  Astronomical Society, 449, 3559, \dodoi{10.1093/mnras/stv582}

\bibitem[{Howitt {et~al.}(2018)Howitt, Melatos, \& Delaigle}]{Howitt2018}
Howitt, G., Melatos, A., \& Delaigle, A. 2018, The Astrophysical Journal, 867,
  60, \dodoi{10.3847/1538-4357/aae20a}

\bibitem[{Howitt {et~al.}(2020)Howitt, Melatos, \& Haskell}]{Howitt2020}
Howitt, G., Melatos, A., \& Haskell, B. 2020, arXiv e-prints, 2008,
  arXiv:2008.00365

\bibitem[{Jankowski {et~al.}(2019)}]{Jankowski2019}
Jankowski, F., {et~al.} 2019, Monthly Notices of the Royal Astronomical
  Society, 484, 3691, \dodoi{10.1093/mnras/sty3390}

\bibitem[{Janssen \& Stappers(2006)}]{Janssen2006}
Janssen, G.~H., \& Stappers, B.~W. 2006, A\&A, 457, 611,
  \dodoi{10.1051/0004-6361:20065267}

\bibitem[{Jensen(1998)}]{Jensen1998}
Jensen, H.~J. 1998, Self-{{Organized Criticality}}. {{Emergent Complex
  Behavior}} in {{Physical}} and {{Biological Systems}}, Cambridge {{Lecture
  Notes}} in {{Physics}} ({Cambridge}: {Cambridge University Press})

\bibitem[{Jones(1998)}]{Jones1998}
Jones, P.~B. 1998, Monthly Notices of the Royal Astronomical Society, 296, 217,
  \dodoi{10.1046/j.1365-8711.1998.01464.x}

\bibitem[{Kagan \& Jackson(1991)}]{Kagan1991}
Kagan, Y.~Y., \& Jackson, D.~D. 1991, Geophysical Journal International, 104,
  117, \dodoi{10.1111/j.1365-246X.1991.tb02498.x}

\bibitem[{Khomenko \& Haskell(2018)}]{Khomenko2018}
Khomenko, V., \& Haskell, B. 2018, Publications of the Astronomical Society of
  Australia, 35, e020, \dodoi{10.1017/pasa.2018.12}

\bibitem[{Kingman(1993)}]{Kingman1993}
Kingman, J. F.~C. 1993, Poisson Processes ({Oxford: Clarendon Press; New York:
  Oxford University Press, 1993.})

\bibitem[{Larson \& Link(2002)}]{Larson2002}
Larson, M.~B., \& Link, B. 2002, Monthly Notices of the Royal Astronomical
  Society, 333, 613, \dodoi{10.1046/j.1365-8711.2002.05439.x}

\bibitem[{Link {et~al.}(1999)Link, Epstein, \& Lattimer}]{Link1999a}
Link, B., Epstein, R.~I., \& Lattimer, J.~M. 1999, Physical Review Letters, 83,
  3362, \dodoi{10.1103/PhysRevLett.83.3362}

\bibitem[{Link \& Epstein(1991)}]{Link1991}
Link, B.~K., \& Epstein, R.~I. 1991, Astrophysical Journal, 373, 592

\bibitem[{L{\"o}nnborn {et~al.}(2019)L{\"o}nnborn, Melatos, \&
  Haskell}]{Lonnborn2019}
L{\"o}nnborn, J.~R., Melatos, A., \& Haskell, B. 2019, Monthly Notices of the
  Royal Astronomical Society, 487, 702, \dodoi{10.1093/mnras/stz1302}

\bibitem[{Lower {et~al.}(2020)}]{Lower2020}
Lower, M.~E., {et~al.} 2020, arXiv:2002.12481 [astro-ph].
\newblock \doarXiv{2002.12481}

\bibitem[{Lyne {et~al.}(2000)Lyne, Shemar, \& Smith}]{Lyne2000a}
Lyne, A.~G., Shemar, S.~L., \& Smith, F.~G. 2000, Monthly Notices of the Royal
  Astronomical Society, 315, 534, \dodoi{10.1046/j.1365-8711.2000.03415.x}

\bibitem[{Manchester {et~al.}(2005)}]{Manchester2005}
Manchester, R.~N., {et~al.} 2005, The Astronomical Journal, 129, 1993,
  \dodoi{10.1109/URSIGASS.2014.6929987}

\bibitem[{Mastrano \& Melatos(2005)}]{Mastrano2005}
Mastrano, A., \& Melatos, A. 2005, Monthly Notices of the Royal Astronomical
  Society, 361, 927, \dodoi{10.1111/j.1365-2966.2005.09219.x}

\bibitem[{McCulloch {et~al.}(1990)}]{McCulloch1990}
McCulloch, P.~M., {et~al.} 1990, Nature, 346, 822, \dodoi{10.1038/346822a0}

\bibitem[{Melatos {et~al.}(2015)Melatos, Douglass, \& Simula}]{Melatos2015}
Melatos, A., Douglass, J.~A., \& Simula, T.~P. 2015, The Astrophysical Journal,
  807, 132, \dodoi{10.1088/0004-637X/807/2/132}

\bibitem[{Melatos \& Drummond(2019)}]{Melatos2019}
Melatos, A., \& Drummond, L.~V. 2019, ApJ, 885, 37,
  \dodoi{10.3847/1538-4357/ab44c3}

\bibitem[{Melatos {et~al.}(2018)Melatos, Howitt, \& Fulgenzi}]{Melatos2018}
Melatos, A., Howitt, G., \& Fulgenzi, W. 2018, ApJ, 863, 196,
  \dodoi{10.3847/1538-4357/aad228}

\bibitem[{Melatos {et~al.}(2008)Melatos, Peralta, \& Wyithe}]{Melatos2008}
Melatos, A., Peralta, C., \& Wyithe, J. S.~B. 2008, The Astrophysical Journal,
  672, 1103, \dodoi{10.1086/523349}

\bibitem[{Melatos \& Warszawski(2009)}]{Melatos2009}
Melatos, A., \& Warszawski, L. 2009, The Astrophysical Journal, 700, 1524,
  \dodoi{10.1088/0004-637X/700/2/1524}

\bibitem[{Melatos {et~al.}(2020)}]{Melatos2020hmm}
Melatos, A., {et~al.} 2020, The Astrophysical Journal, 896, 78,
  \dodoi{10.3847/1538-4357/ab9178}

\bibitem[{Middleditch {et~al.}(2006)}]{Middleditch2006}
Middleditch, J., {et~al.} 2006, The Astrophysical Journal, 652, 1531,
  \dodoi{10.1086/508736}

\bibitem[{Mongiov{\`i} {et~al.}(2017)Mongiov{\`i}, Russo, \&
  Sciacca}]{Mongiovi2017}
Mongiov{\`i}, M.~S., Russo, F.~G., \& Sciacca, M. 2017, Monthly Notices of the
  Royal Astronomical Society, 469, 2141, \dodoi{10.1093/mnras/stx827}

\bibitem[{Newman(1996)}]{Newman1996a}
Newman, M. E.~J. 1996, Proceedings of the Royal Society of London Series B,
  263, 1605

\bibitem[{Newman \& Sneppen(1996)}]{Newman1996}
Newman, M. E.~J., \& Sneppen, K. 1996, Phys. Rev. E, 54, 6226,
  \dodoi{10.1103/PhysRevE.54.6226}

\bibitem[{Palfreyman {et~al.}(2018)}]{Palfreyman2018}
Palfreyman, J., {et~al.} 2018, Nature, 556, 219,
  \dodoi{10.1038/s41586-018-0001-x}

\bibitem[{Peralta {et~al.}(2006)}]{Peralta2006}
Peralta, C., {et~al.} 2006, The Astrophysical Journal, 651, 1079,
  \dodoi{10.1086/507576}

\bibitem[{Pizzochero(2011)}]{Pizzochero2011}
Pizzochero, P.~M. 2011, ApJL, 743, L20, \dodoi{10.1088/2041-8205/743/1/L20}

\bibitem[{Pizzochero {et~al.}(2020)Pizzochero, Montoli, \&
  Antonelli}]{Pizzochero2020}
Pizzochero, P.~M., Montoli, A., \& Antonelli, M. 2020, Astronomy and
  Astrophysics, 636, A101, \dodoi{10.1051/0004-6361/201937019}

\bibitem[{Press {et~al.}(2007)}]{Press2007}
Press, W.~H., {et~al.} 2007, Numerical {{Recipes}} 3rd {{Edition}}: {{The Art}}
  of {{Scientific Computing}} ({Cambridge University Press})

\bibitem[{Ruderman {et~al.}(1998)Ruderman, Zhu, \& Chen}]{Ruderman1998}
Ruderman, M., Zhu, T., \& Chen, K. 1998, The Astrophysical Journal, 492, 267,
  \dodoi{10.1086/305026}

\bibitem[{Shannon {et~al.}(2016)}]{Shannon2016}
Shannon, R.~M., {et~al.} 2016, Monthly Notices of the Royal Astronomical
  Society, 459, 3104, \dodoi{10.1093/mnras/stw842}

\bibitem[{Sneppen \& Newman(1997)}]{Sneppen1997}
Sneppen, K., \& Newman, M. 1997, Physica D: Nonlinear Phenomena, 110, 209,
  \dodoi{10.1016/S0167-2789(97)00128-0}

\bibitem[{Srinivasan {et~al.}(1990)}]{Srinivasan1990}
Srinivasan, G., {et~al.} 1990, Current Science, 59, 31

\bibitem[{Stappers {et~al.}(2011)}]{Stappers2011}
Stappers, B.~W., {et~al.} 2011, Astronomy and Astrophysics, 530, A80,
  \dodoi{10.1051/0004-6361/201116681}

\bibitem[{Utsu {et~al.}(1995)}]{Utsu1995}
Utsu, T., {et~al.} 1995, Journal of Physics of the Earth, 43, 1,
  \dodoi{10.4294/jpe1952.43.1}

\bibitem[{{van der Hoek} \& Elliott(2018)}]{VanDerHoek2018}
{van der Hoek}, J., \& Elliott, R.~J. 2018, Introduction to {{Hidden
  Semi}}-{{Markov Models}}, London {{Mathematical Society Lecture Note Series}}
  ({Cambridge}: {Cambridge University Press}), \dodoi{10.1017/9781108377423}

\bibitem[{Warszawski \& Melatos(2011)}]{Warszawski2011}
Warszawski, L., \& Melatos, A. 2011, Monthly Notices of the Royal Astronomical
  Society, 415, 1611, \dodoi{10.1111/j.1365-2966.2011.18803.x}

\bibitem[{Warszawski \& Melatos(2013)}]{Warszawski2013}
---. 2013, Monthly Notices of the Royal Astronomical Society, 428, 1911,
  \dodoi{10.1093/mnras/sts108}

\bibitem[{Warszawski {et~al.}(2012)Warszawski, Melatos, \&
  Berloff}]{Warszawski2012}
Warszawski, L., Melatos, A., \& Berloff, N.~G. 2012, Physical Review B -
  Condensed Matter and Materials Physics, 85,
  \dodoi{10.1103/PhysRevB.85.104503}

\bibitem[{Yu \& Liu(2017)}]{Yu2017b}
Yu, M., \& Liu, Q.-J. 2017, Monthly Notices of the Royal Astronomical Society,
  468, 3031, \dodoi{10.1093/mnras/stx702}

\end{thebibliography}

\end{document}